\documentclass[myepj-spec]{mySvjour}
\usepackage{graphicx}
\usepackage[breaklinks]{hyperref}
\usepackage{color}
\usepackage{xspace}
\usepackage[square,sort&compress,numbers]{natbib}   
\usepackage{verbatim}
\usepackage{amsmath,amssymb,amsfonts} 
\usepackage{tabularx}
\usepackage{lmodern,bm}                
\usepackage[T1]{sansmath} 
\SetMathAlphabet{\mathsfbf}{sans}{\sansmathencoding}{\sfdefault}{bx}{sl}
\usepackage{etoolbox}
\AtBeginEnvironment{sansmath}{}{}{}

\newcommand{\figsize}{12.5cm}
\newcommand{\plotsize}{8.0cm}
\newcommand{\fighspace}{0.4cm}

\definecolor{darkblue1}{rgb}{0,0,.2}
\definecolor{darkblue}{rgb}{0,0,.2}
\definecolor{darkred}{rgb}{0.5,0,0}
\pagecolor{white} 
\color{black}     
\hypersetup{breaklinks=true, 
            colorlinks=true, 
            linkcolor=darkblue1, 
            menucolor=darkblue1, 
            urlcolor=darkblue1,
            citecolor=darkblue1,
            pdftitle={},
            pdfauthor={},
            pdfsubject={},
            pdfkeywords={},
            pdfproducer={}
}
\bibstyle{plain}

\RequirePackage{xspace}

\newcommand{\beq}{\begin{equation}}
\newcommand{\eeq}{\end{equation}}
\newcommand{\beqn}{\begin{eqnarray}}
\newcommand{\eeqn}{\end{eqnarray}}
\newcommand{\beqns}{\begin{eqnarray*}}
\newcommand{\eeqns}{\end{eqnarray*}}
\newcommand{\bei}{\begin{itemize}}
\newcommand{\eei}{\end{itemize}}

\newcommand{\rar}{\rightarrow}

\newcommand{\mev}{\ensuremath{\mathrm{\;Me\kern -0.1em V}}\xspace}
\newcommand{\gev}{\ensuremath{\mathrm{\;Ge\kern -0.1em V}}\xspace}

\usepackage{relsize}
\def\babar{\mbox{\slshape B\kern-0.1em{\smaller A}\kern-0.1em
    B\kern-0.1em{\smaller A\kern-0.2em R}}}
\def\cmd3{\mbox{CMD-3}}
\newcommand{\pp}{\ensuremath{\pi^+\pi^-}\xspace}

\newcommand{\eetopp}{\ensuremath{e^+e^-\!\rightarrow\pi^+\pi^-}\xspace}
\def\amu{$a_\mu$}

\begin{document}

\title{\boldmath Tensions in $e^+e^-\to\pi^+\pi^-(\gamma)$ measurements: the new landscape of data-driven hadronic vacuum polarization predictions for the muon $g$\,--\,2}

\author{Michel Davier\inst{1} \and Andreas Hoecker\inst{2} \and Anne-Marie Lutz\inst{1} \and Bogdan Malaescu\inst{3} \and Zhiqing Zhang\inst{1}}
 
\institute{IJCLab, Universit\'e Paris-Saclay et CNRS/IN2P3, Orsay, France \and
CERN, CH--1211, Geneva 23, Switzerland \and
LPNHE, Sorbonne Université, Université Paris Cité, CNRS/IN2P3, Paris, France }

\twocolumn[{%
  \begin{@twocolumnfalse}

    \begin{flushright}
      \normalsize
      \today
    \end{flushright}

    \vspace{-2cm}

\abstract{
The situation of the experimental data used in the dispersive evaluation of the hadronic vacuum polarization contribution to the anomalous magnetic moment of the muon is assessed in view of two recent measurements: $e^+e^- \to \pi^+\pi^-$ cross sections in the $\rho$ resonance region by \cmd3\ and a study of higher-order radiative effects in the initial-state-radiation processes $e^+e^- \to \mu^+\mu^-\gamma$ and $e^+e^- \to \pi^+\pi^-\gamma$ by BABAR. The impact of the latter study on the KLOE and BESIII cross-section measurements is evaluated and found to be indicative of 
larger systematic effects than uncertainties assigned. The new situation also warrants a reappraisal of the independent information provided by hadronic $\tau$ decays, including state-of-the-art isospin-breaking corrections. The findings cast a new light on the longstanding deviation between the muon $g$\,--\,2 measurement and the Standard Model prediction using the data-driven dispersive approach, and the comparison with lattice QCD calculations.} 
\maketitle

  \end{@twocolumnfalse}
}]

\section{Introduction} 

\sloppy
The muon anomalous magnetic moment, characterized by the gyromagnetic anomaly, $a_\mu=(g-2)/2$, is the subject of notable current interest. Recent measurements at Fermilab~\cite{Muong-2:2023cdq,Muong-2:2021ojo} with increased precision exhibit a spectacular excess compared to the Standard Model (SM) prediction~\cite{Aoyama:2020ynm} 
whose lowest-order (LO) hadronic vacuum  polarization~(HVP) contribution is evaluated with dispersion integrals involving $e^+e^-\to \mathrm{hadrons}$ cross-section data~\cite{Davier:2017zfy,Keshavarzi:2018mgv,Colangelo:2018mtw,Hoferichter:2019mqg,Davier:2019can,Keshavarzi:2019abf}. 
Other contributions are from quantum electrodynamics~\cite{aoyama:2012wk,Aoyama:2019ryr}, electroweak interactions~\cite{czarnecki:2002nt,gnendiger:2013pva}, NLO HVP~\cite{Keshavarzi:2019abf}, NNLO HVP~\cite{kurz:2014wya}, hadronic light-by-light~\cite{melnikov:2003xd,masjuan:2017tvw,Colangelo:2017fiz,hoferichter:2018kwz,gerardin:2019vio,bijnens:2019ghy,colangelo:2019uex,pauk:2014rta,danilkin:2016hnh,jegerlehner:2017gek,knecht:2018sci,eichmann:2019bqf,roig:2019reh,Blum:2019ugy,colangelo:2014qya}.
While the observed $5\sigma$ deviation could be regarded as a serious clue for physics beyond the SM, it must be taken with extreme caution in view of significant tensions among the data sets entering the HVP calculations. Although the decade-long discrepancy between the two most precise results of the $e^+e^-\rar\pi^+\pi^-(\gamma)$ cross section\footnote{If not explicitly stated, final state photon radiation is implied throughout this paper for all hadronic final states.} from KLOE~\cite{KLOE:2008fmq,KLOE:2010qei,KLOE:2012anl,KLOE-2:2017fda} and BABAR~\cite{BaBar:2009wpw,BaBar:2012bdw} was already taken into account in the systematic uncertainty assigned to the prediction~\cite{Davier:2019can,Aoyama:2020ynm}, the recent measurement of the same process by \cmd3~\cite{CMD-3:2023alj} is in conflict with all previous determinations, thus requiring a close scrutiny of all the $e^+e^-$ input data.

A tension of a different nature arose almost four years ago with the first precise HVP calculation using QCD on the lattice~\cite{Borsanyi:2020mff} that resulted in a $2.1\sigma$ larger lowest-order contribution than the dispersive analysis. The tension is exacerbated to $3.7\sigma$ if the comparison is  restricted to an intermediate HVP window in Euclidean time~\cite{RBC:2018dos}, which can be calculated more precisely on the lattice. Confirmation of this discrepancy has since then been obtained by several independent lattice groups~\cite{Ce:2022kxy,ExtendedTwistedMass:2022jpw,RBC:2023pvn,FermilabLatticeHPQCD:2023jof}. This situation calls again for specific studies cross-checking both approaches~\cite{Colangelo:2022vok,Davier:2023cyp}. 

This paper reviews the existing tensions among the \eetopp cross-section measurements, and discusses in detail systematic uncertainties related to higher-order (HO) effects~\cite{BaBar:2023xiy} in the measurements relying on initial state photon radiation. In view of the results obtained we reappraise the use of $\tau$ hadronic spectral functions in the dispersive approach, and discuss the discrepancies of the dispersive HVP calculations with lattice QCD and the $a_\mu$ experimental result.

\section{Tensions among the $e^+e^-\to\pi^+\pi^-(\gamma)$ data sets}
\label{sec:tensions-2pi}

The \eetopp channel contributes with 73\% to the lowest-order HVP contribution to $a_\mu$ in the dispersive approach, and 58\% to its  uncertainty-squared. It also leads to the largest observed discrepancies among some of the most precise data sets. The studies in this paper therefore focus on that process. 

The longest known and most critical  tensions occur between precise cross section measurements from KLOE and BABAR. Albeit heavily discussed in the framework of the Muon $g$\,--\,2 Theory Initiative~\cite{g-2-TI}, no understanding of the difference could be achieved and consequently no solution to the problem emerged. The discrepancy was bridged by inflated uncertainties in the corresponding HVP contribution.  

\begin{figure}[t]
\begin{center}
\includegraphics[width=\plotsize]{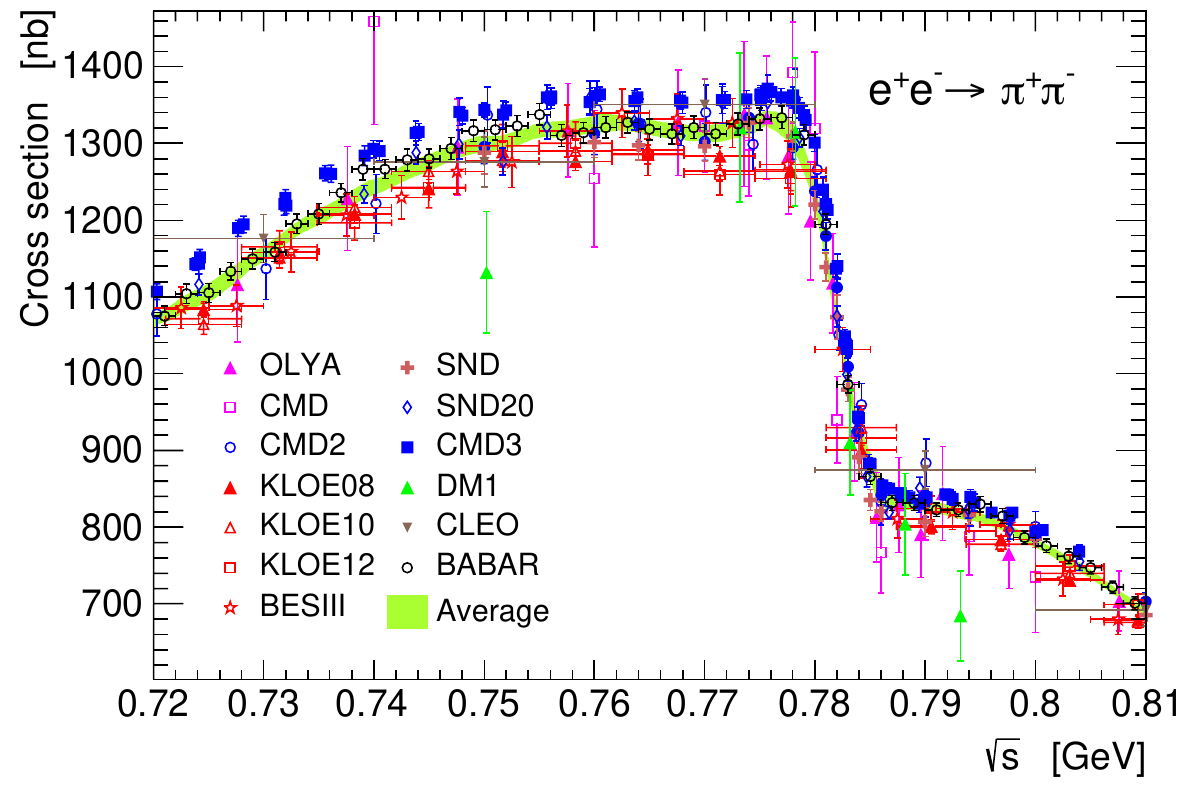}
\end{center}
\vspace{-0.2cm}
\caption[.]{ Bare \eetopp cross section versus centre-of-mass energy in the $\rho$ peak region. The error bars of the data points include statistical and systematic uncertainties added in quadrature. The green band shows the HVPTools combination within its $1\sigma$ uncertainty. }
\label{fig:pipiall}
\end{figure}

The available \eetopp cross-section measurements, zoomed into the $\rho$ peak region, are shown in Fig.~\ref{fig:pipiall}. 
Their combination and $1\sigma$ uncertainty, obtained using the DHMZ methodology implemented in the HVPTools software~\cite{Davier:2010rnx,Davier:2010fmf}, is indicated by the green band.
The spline-based combination procedure\footnote{Since the main purpose of the combination here is to provide a common reference for comparing the various measurements, we do not employ the analyticity-based constraints used in Ref.~\cite{Davier:2019can}.} takes into account all known correlations and accounts for measurement tensions. It has been thoroughly validated through closure tests~\cite{Davier:2010rnx}.
Compared to our last update~\cite{Davier:2019can}, we added the more recent SND20~\cite{SND:2020nwa} and \cmd3~\cite{CMD-3:2023alj} data, while also employing an updated version of the covariance matrix provided by  BESIII~\cite{BESIII:2015equ}.

\begin{figure*}[htbp]
\begin{center}
\includegraphics[width=\plotsize]{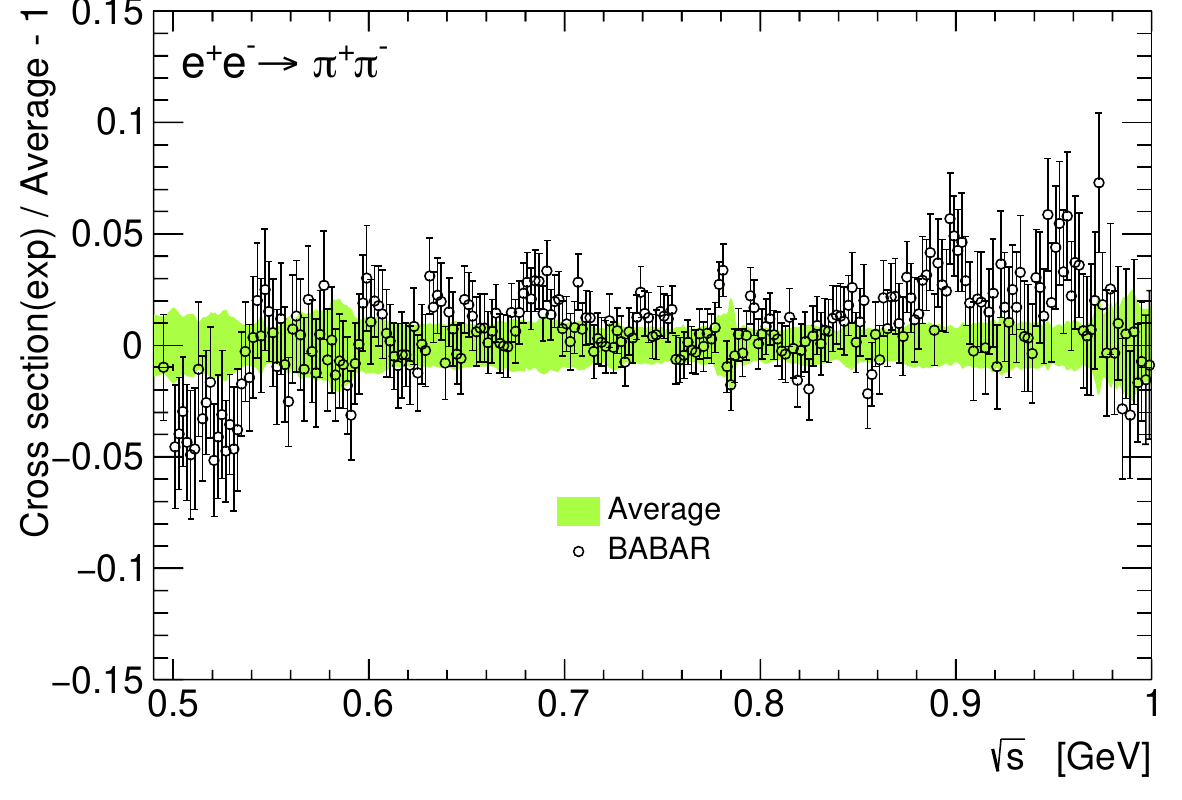}\hspace{\fighspace}
\includegraphics[width=\plotsize]{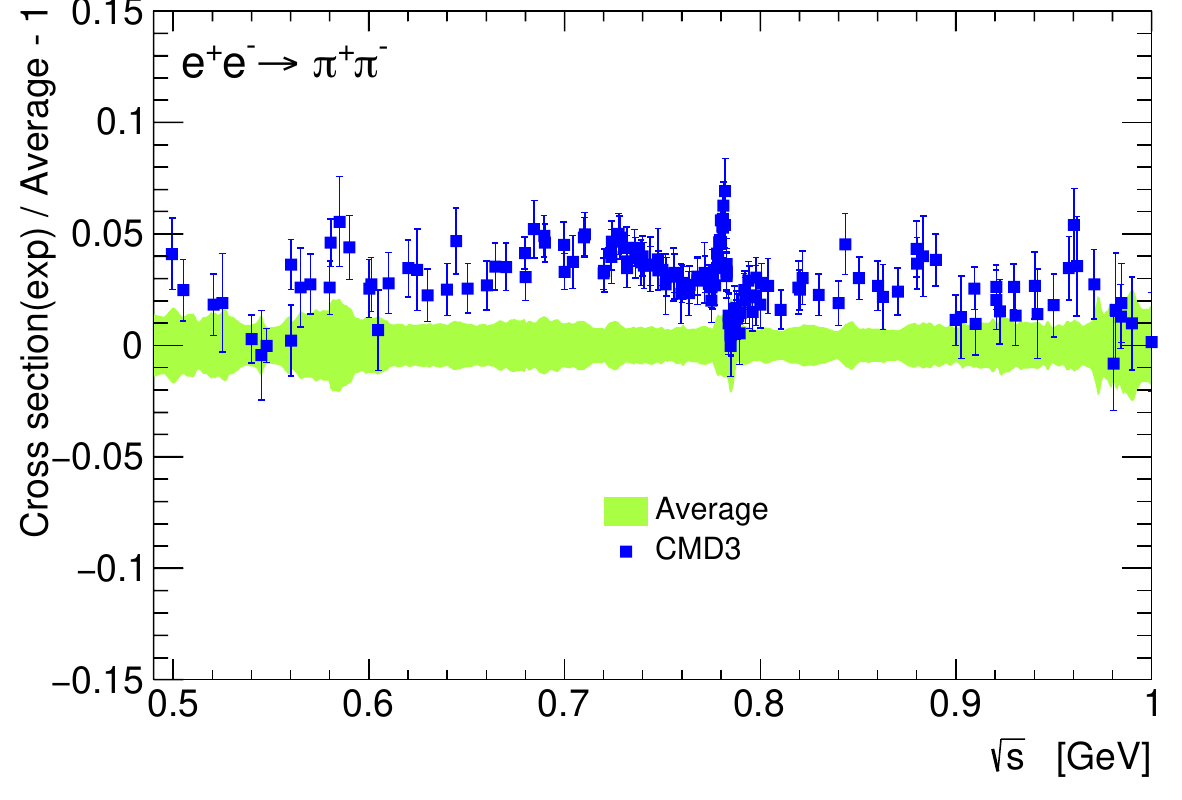}
\vspace{0.2cm}

\includegraphics[width=\plotsize]{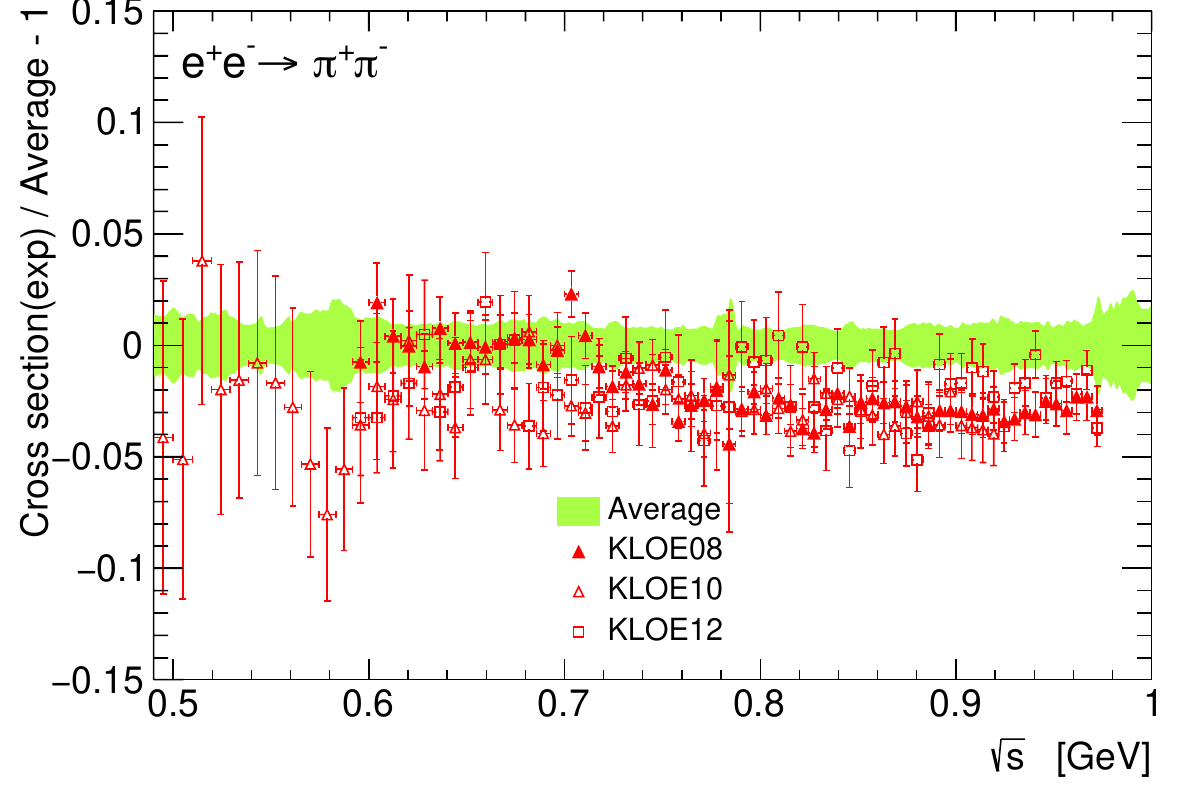}\hspace{\fighspace}
\includegraphics[width=\plotsize]{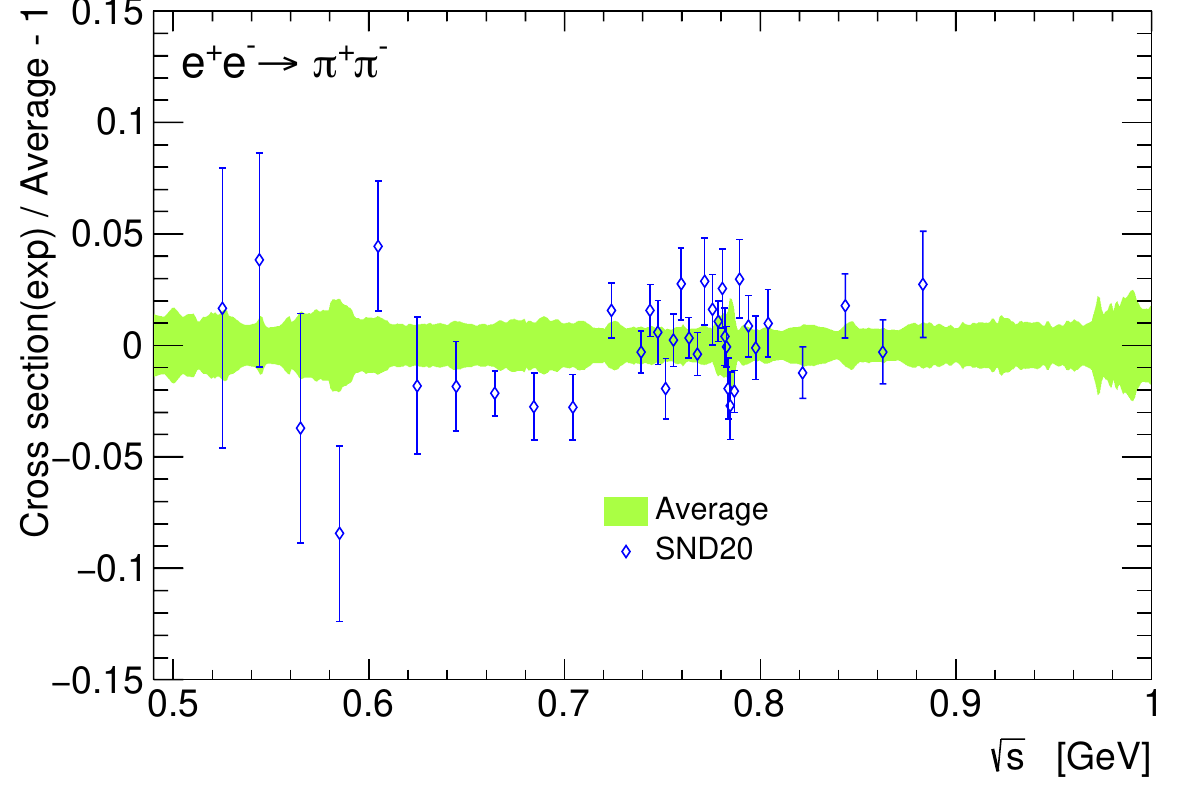}
\vspace{0.2cm}

\includegraphics[width=\plotsize]{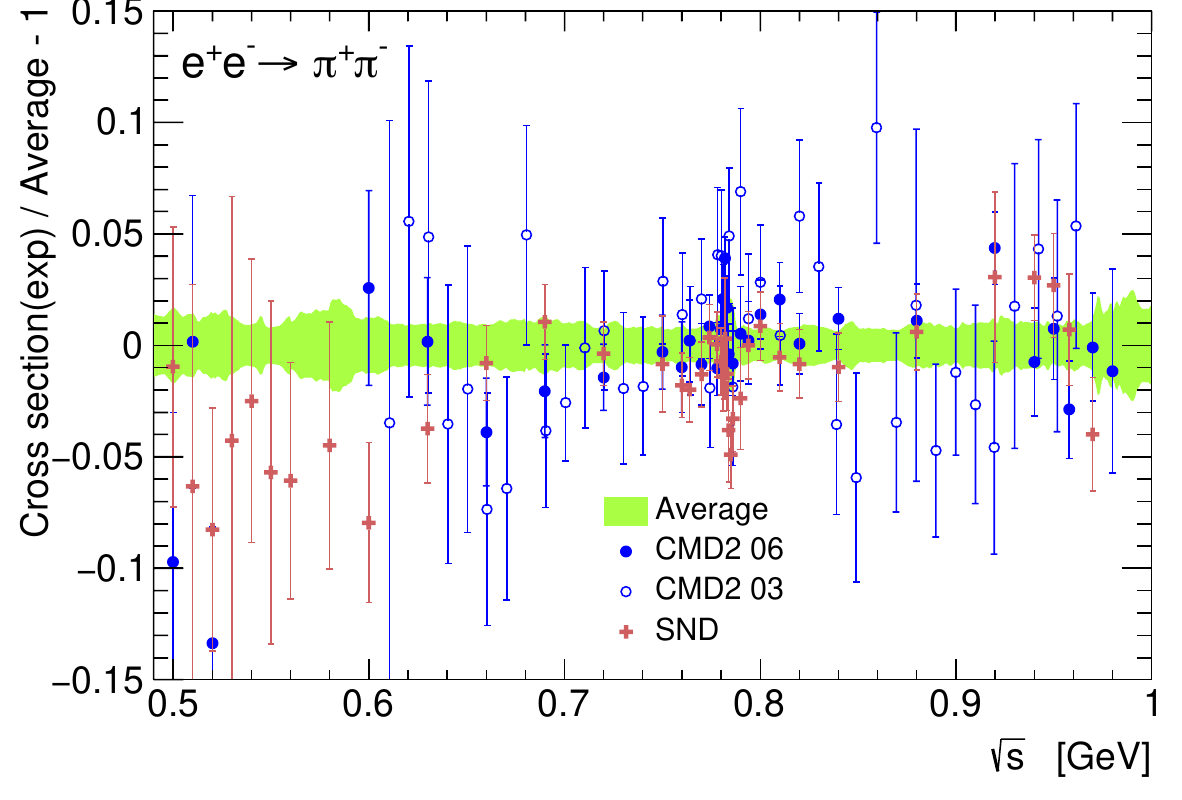}\hspace{\fighspace}
\includegraphics[width=\plotsize]{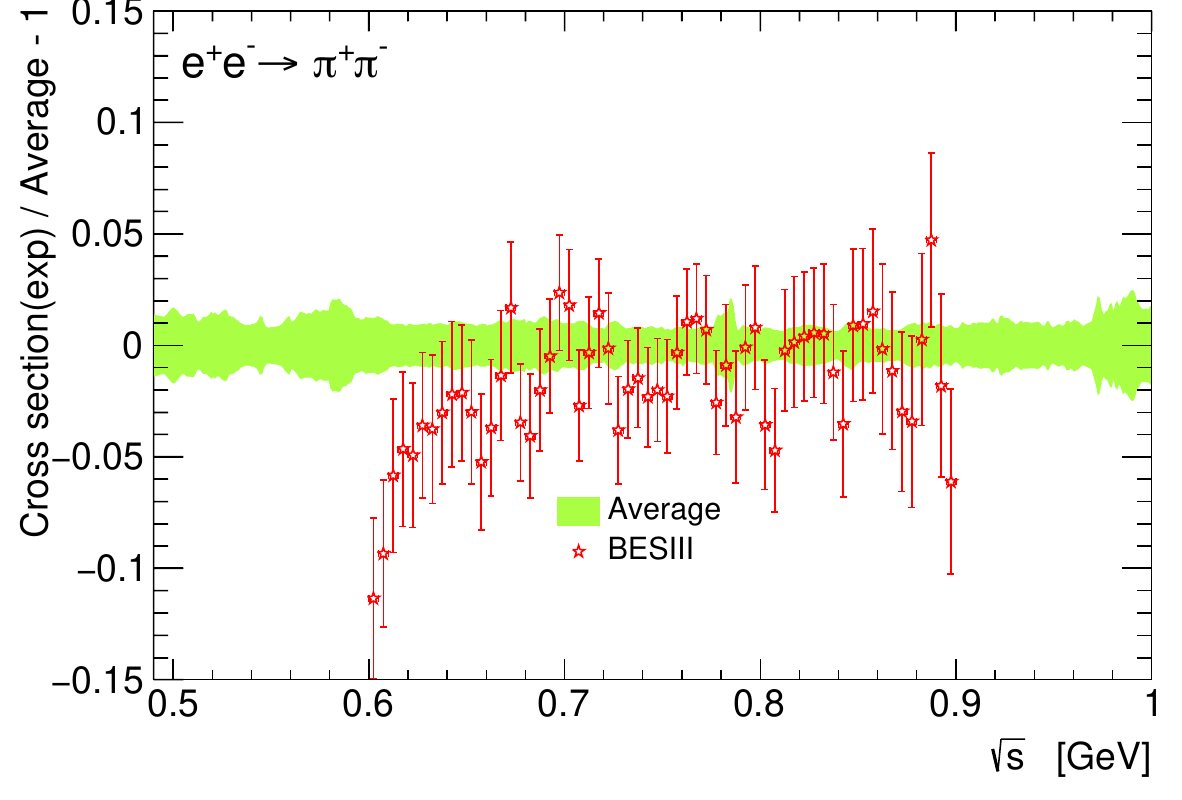}
\end{center}
\vspace{-0.2cm}
\caption[.]{ Comparison between  \eetopp cross-section measurements from BABAR~\cite{BaBar:2009wpw,BaBar:2012bdw}, KLOE\,08~\cite{KLOE:2008fmq}, KLOE\,10~\cite{KLOE:2010qei}, KLOE\,12~\cite{KLOE:2012anl}, BESIII~\cite{BESIII:2015equ}, 
CMD-2\,03~\cite{CMD-2:2003gqi}, CMD-2\,06~\cite{CMD-2:2006gxt}, SND~\cite{Achasov:2006vp}, SND20~\cite{SND:2020nwa}, \cmd3~\cite{CMD-3:2023alj}, and the HVPTools combination. The error bars include statistical and systematic uncertainties added in quadrature. 
}
\label{fig:comppipiZoomRho}
\end{figure*}

\begin{figure}[htbp]
\begin{center}
\includegraphics[width=\plotsize]{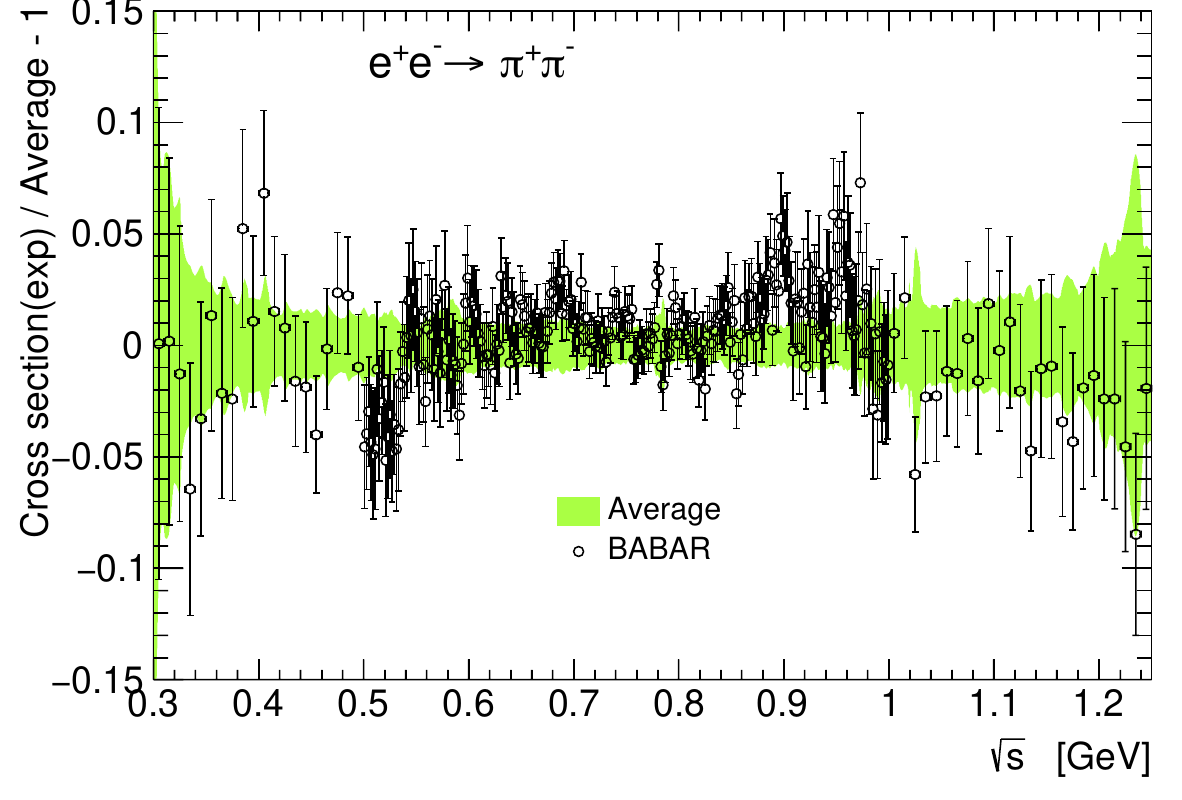} \\ 
\includegraphics[width=\plotsize]{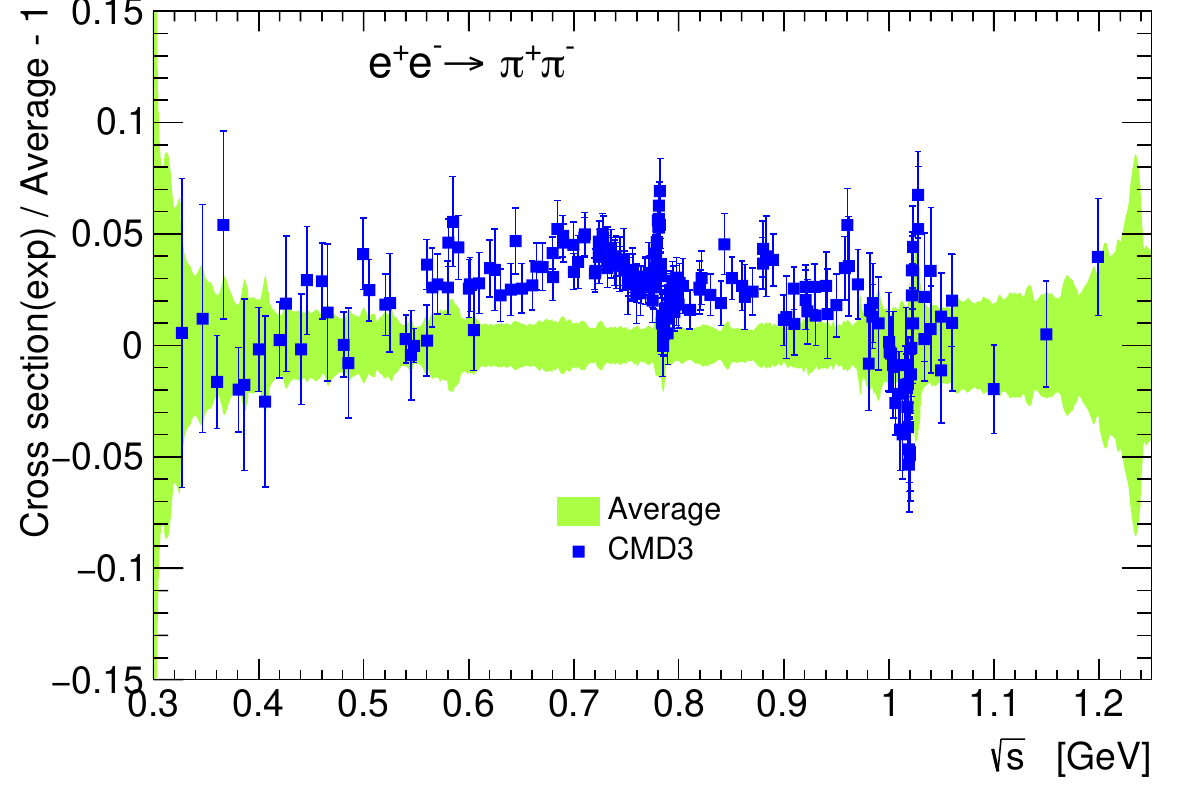}
\end{center}
\vspace{-0.2cm}
\caption[.]{ Comparison between \eetopp cross-section measurements from BABAR~\cite{BaBar:2009wpw,BaBar:2012bdw} (top panel), \cmd3~\cite{CMD-3:2023alj} (bottom), and the HVPTools combination of all available data in the $\sqrt{s}$ range covered by \cmd3. The error bars include statistical and systematic uncertainties added in quadrature. 
}
\label{fig:comppipiZoomOut}
\end{figure}

Relative comparisons between the most precise individual measurements and the combination are shown for the $\rho$ resonance region in Fig.~\ref{fig:comppipiZoomRho}, and for the BABAR and \cmd3 data in a wider window in Fig.~\ref{fig:comppipiZoomOut}.
A large tension arises between \cmd3 and KLOE, which provide the, respectively, largest and  smallest cross-section measurements. Tensions are also observed between BABAR and \cmd3 in the central $\rho$ resonance region, while they agree at low and high energies.
The  \cmd3 data also exhibit a   
2.8$\sigma$ discrepancy with the older CMD-2 results by the same collaboration~\cite{CMD-2:2006gxt}.
Extensive discussions with \cmd3/2 physicists in the framework of the Muon $g$\,--\,2 Theory Initiative~\cite{cmd3-discuss} did not reveal any obvious problem in the new results. A summary of these discussions is available~\cite{cmd3-review-summary}.

\begin{figure}[htbp]
\begin{center}
\includegraphics[width=\plotsize]{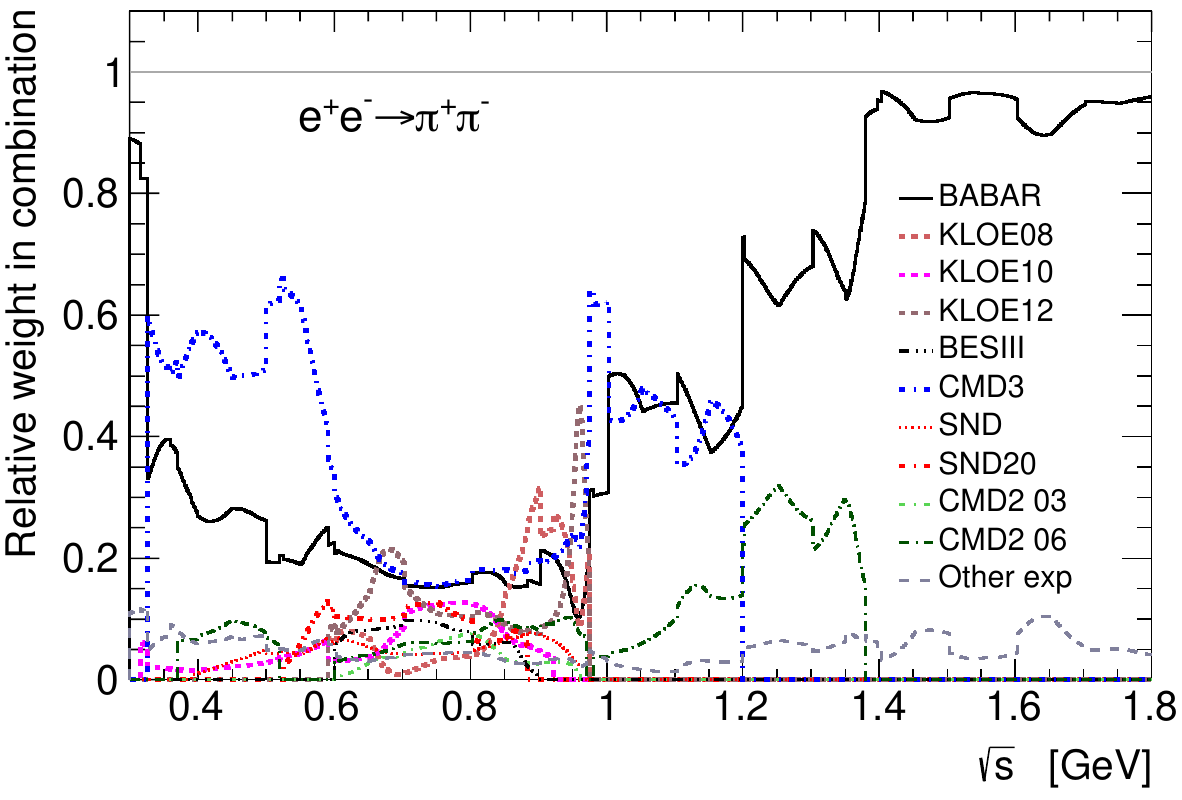} \\ 
\includegraphics[width=\plotsize]{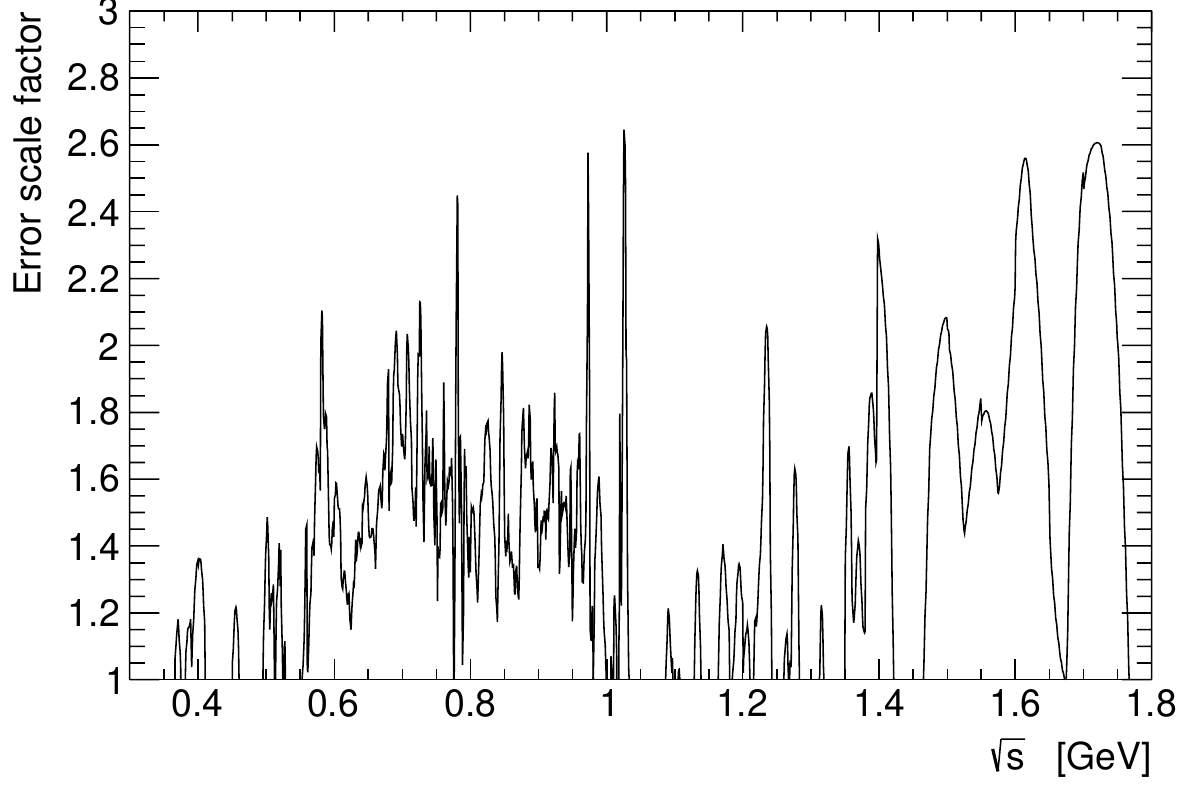}
\end{center}
\vspace{-0.2cm}
\caption[.]{ Top: relative local weight per measurement contributing to the \eetopp cross-section combination versus centre-of-mass energy. Bottom: local uncertainty scale factor versus centre-of-mass energy applied to the combined \pp cross-section uncertainty to account for inconsistencies among the measurements. }
\label{fig:WeightsChi2}
\end{figure}

Figure~\ref{fig:WeightsChi2} (top) shows the local combination weights versus $\sqrt{s}$ for each data set.
They take into account the uncertainties of the measurements and their correlations, as well as the corresponding point-spacing and binning~\cite{Davier:2010fmf,Davier:2010rnx}.
While previously
the BABAR and KLOE measurements dominated the combination over the entire energy range, the more recent \cmd3 and SND20 data receive important weights, too.
The group of experiments labelled ``Other exp'' corresponds to older data, often with incomplete radiative corrections, which receive small weights throughout.

The bottom panel of Fig.~\ref{fig:WeightsChi2} displays the uncertainty scale factor versus $\sqrt{s}$, derived based on the local compatibility among the measurements~\cite{Davier:2010fmf,Davier:2010rnx}.\footnote{While the uncertainty rescaling is applied to the combined \pp cross-section uncertainty to account for local inconsistencies among the measurements, a global systematic tension must also be taken into account in the HVP calculation~\cite{Davier:2019can}.} 
Large scale factors due to tensions indicate the presence of systematic effects that are not included in the measurement uncertainties. They require a conservative uncertainty treatment in the combination~\cite{Davier:2019can,Aoyama:2020ynm}.

\begin{figure}[htbp]
\begin{center}
\includegraphics[width=\plotsize]{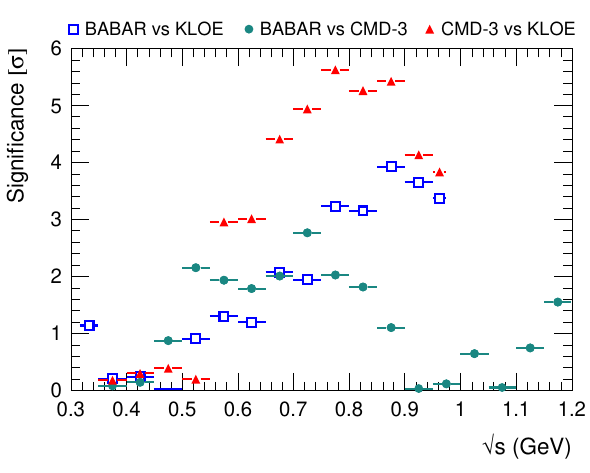} \\ 
\includegraphics[width=\plotsize]{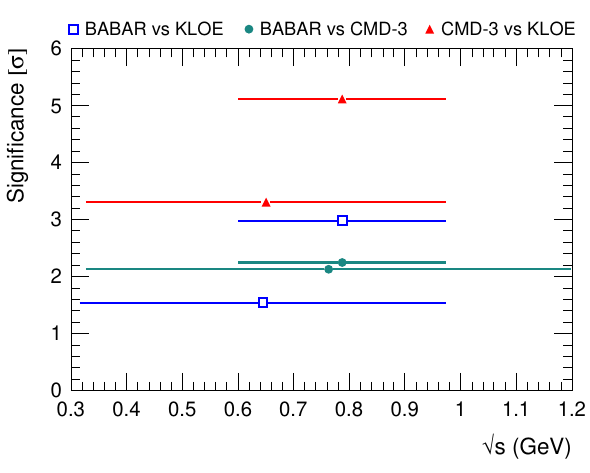}
\end{center}
\vspace{-0.3cm}
\caption[.]{Significance of the difference between pairs of the three most precise \eetopp experiments for narrow energy intervals of $50\mev$ or less (top) and larger energy intervals (bottom) indicated by the horizontal lines.}
\label{fig:sigma}
\end{figure}

Figure~\ref{fig:sigma} shows the pull magnitude (significance) between pairs of the three most precise \eetopp experiments, computed as the absolute value of the difference  of the contributions to \amu\ divided by its  uncertainty, in various energy intervals. The three KLOE measurements~\cite{KLOE:2008fmq, KLOE:2010qei,KLOE:2012anl} have been combined into one data set~\cite{KLOE-2:2017fda}. 
The difference between BABAR and \cmd3 rises to a significance of 2--3$\sigma$ on the $\rho$ peak, while reasonable agreement is seen at lower and higher energies.
The differences between BABAR and KLOE are also at the 2--3$\sigma$ level in the $\rho$ peak region, reaching up to $4\sigma$ at higher energy, while good agreement is seen at lower energy.
The largest differences are observed between \cmd3 and KLOE, with significance above $5\sigma$ around the $\rho$ peak.
When probing the broader energy interval 0.6--0.975$\gev$, covering the $\rho$ peak,  the significance of the difference between BABAR and \cmd3 is $2.2\sigma$, that between BABAR and KLOE is $3.0\sigma$, while \cmd3 and KLOE differ by $5.1\sigma$ (Fig.~\ref{fig:sigma}, bottom).
When extending the comparisons to the maximal regions of overlap between pairs of experiments, the differences are diluted to $2.1\sigma$ between BABAR and \cmd3, $1.5\sigma$ between BABAR and KLOE, and $3.3\sigma$ between \cmd3\ and KLOE, respectively, owing to the better inter-experiment agreement and larger KLOE uncertainties below and above the peak of the resonance.

The interesting possibility to resolve the tensions between different data sets using basic theoretical constraints on the pion form factor from analyticity and unitary has been investigated~\cite{Colangelo:2020lcg}. However, the theory-constrained fits are loose enough to accommodate even the extreme cases of KLOE and CMD-3~\cite{Colangelo:2023rqr}.

\section{BABAR study of additional photon radiation}
\label{sec:BABAR_NLO_NNLO}

The BABAR collaboration performed unique measurements of additional photon radiation in the initial state radiation (ISR) processes $e^+e^-\to\mu^+\mu^-\gamma$ and $e^+e^-\to\pi^+\pi^-\gamma$. Hard NLO radiation with one additional photon was studied in Refs.~\cite{BaBar:2009wpw,BaBar:2012bdw}. A new analysis~\cite{BaBar:2023xiy} based on the full available data set extended that study and included for the first time the measurement of hard NNLO processes with two additional photons from either initial or final state radiation (FSR). 
The paper also includes comparisons with  predictions from the NLO \textsc{Phokhara} and the partial NNLO \textsc{AfkQed}~\cite{afkqed} Monte Carlo generators. 

In the following we use the notation LO, NLO, NNLO to specify the true QED order defined with respect to the lowest-order ISR process, while the same symbols taken within quotes, `LO', `NLO', `NNLO', refer to reconstructed topologies with various photon multiplicities.
We summarise here the main findings of the BABAR study~\cite{BaBar:2023xiy}:
\begin{itemize}
    \item `NNLO' contributions with additional photon energies above 200\mev (100\mev) for the most (least) energetic one (representing (1.9--3.8)\% of the beam energy in the centre-of-mass frame) are observed in $(3.47\pm0.38)$\% and $(3.36\pm0.39)$\% of the dimuon and dipion events, respectively. These events are dominated by small-angle additional ISR photons. 
    \item The `NLO' 
    event fractions, with one additional detected or kinematically reconstructed photon above 200\mev, predicted by the \textsc{Phokhara} generator exceed the BABAR data, particularly for additional ISR photons at small angle. Over the full measured phase space, including additional ISR and FSR photons, \textsc{Phokhara} over-predicts the hard `NLO' contribution by a factor of $1.25\pm0.05$.
    \item The BABAR cross-section measurements~\cite{BaBar:2009wpw,BaBar:2012bdw} are found to be insensitive to the missing NNLO contributions in (and hard `NLO' excess of)  the \textsc{Phokhara} generator.   
    \item The \textsc{AfkQed} generator approximates real and virtual higher-order (HO) corrections by resumming the leading logarithms.\footnote{\textsc{AfkQed} generates the additional NLO (or NNLO) ISR photons in the collinear approximation using the structure function technique, which is not consistent with the `NLO' angular distribution with respect to the beam axis observed in the BABAR data.
    This shortcoming is not present in the KKMC event generator~\cite{Jadach:1999vf}, which uses a different technique for multi-photon emission, while maintaining the same advantage of leading-log resummation. We have verified that the  `LO', `NLO', and `NNLO' event fractions predicted by AfkQed and KKMC are in agreement with the data.} It provides a reasonable description of the rates and energy distributions of the measured `NLO' and `NNLO' topologies. 
\end{itemize}

\section{Cancellation between soft/virtual and hard photon corrections}
\label{sec:cancellations}

\subsection{Expected behaviour of NLO events and experimental procedures}
\label{sec:cancel-nlo}

To assess the effect of higher-order radiative corrections one needs to evaluate the sensitivity of a given measurement to the presence of additional photon radiation, particularly ISR. 
An analysis that rejects part of the additional ISR requires a compensating correction from an NLO Monte Carlo generator to be consistent, at that order, with the corresponding ISR luminosity computed with the same generator. 
The \textsc{Phokhara} event generator version~9.1~\cite{Campanario:2013uea} incorporates all contributions from NLO QED and thus provides a complete prediction of the ISR process $e^+e^-\to\mu^+\mu^-\gamma(\gamma)$.
It includes the lowest-order (LO) ISR and FSR processes and NLO contributions from real photon emission by the $e^\pm$ beams and the outgoing muons, as well as soft photon emission and  virtual corrections. The sum of the soft and virtual terms is infrared finite and the transition energy between soft and hard emission is chosen within a safe range (5\mev for BABAR simulations) so that both contributions are under control. From an experimental point of view, both LO and soft plus virtual NLO lead to event configurations that are reconstructed in the `LO' topology and kinematics, whereas sufficiently hard NLO radiation necessitates a different kinematic treatment. The lowest energy for NLO photon contributions is experiment dependent. 
In BABAR a value of 50\mev, the energy threshold for a detected photon included in kinematic fits, is representative, although a higher threshold (200\mev) is applied to detected or kinematically reconstructed photons to separate the `NLO' from the `LO' topologies for the final results.

The effects of HO radiative corrections are evaluated using samples of ISR muon-pair events generated with \textsc{Phokhara} in the BABAR conditions: ISR (or FSR) photon at large polar angle ($20^\circ$--$160^\circ$) in the $e^+e^-$ centre-of-mass (CM) system; two-charged-particle mass from threshold to 1.4\gev;  $\sqrt{s}=10.58$\gev CM energy. 
Soft and virtual corrections are studied with the use of samples generated at LO with either ISR only or with ISR and FSR, and samples generated at NLO with either ISR only or the full NLO configuration with ISR, FSR, and their interference. The fraction of hard photon radiation turns out to be rather large because NLO ISR is enhanced by a factor $\ln(s/m_e^2)$. 
It strongly depends on the energy threshold of the additional photon: a fraction of 60\% for $E^\ast_\gamma$ above 5\mev in the centre-of-mass decreases to 38\% above 50\mev and to 25\% above 200\mev. All contributions are dominated by NLO ISR at small angle with respect to the beam axis. 
For example, with 50\mev photon energy threshold the NLO ISR fraction at small angle outside the BABAR acceptance is 27\%, NLO ISR at large angle 8\%, and NLO FSR 3\%. These values illustrate the importance of a thorough understanding and robust correction of effects from HO radiative corrections. 
The situation is very similar for the $e^+e^-\to\pi^+\pi^-\gamma(\gamma)$ ISR process~\footnote{For pions, additional radiation requires a model, usually assuming scalar QED with point-like pions and improved with dispersive methods~\cite{Colangelo:2022lzg}.}.

It is instructive to compare the \textsc{Phokhara} predictions at different orders. For the BABAR conditions the full NLO (LO) cross section for $e^+e^-\rar\mu^+\mu^-\gamma(\gamma)$ amounts to 17.16\;pb (17.45\;pb), a reduction by $-1.7$\% at NLO. Since the NLO cross-section contribution with an additional photon above $50\mev$ corresponds to $38\%\times 17.16/17.45\simeq37$\%, it is almost compensated by a reduction of 39\% due to the soft and virtual contribution. 
This large cancellation between hard and soft/virtual effects is well-known in QED~\cite{KLN}. It requires a careful assessment of the measured and theoretically corrected cross-section fractions.

\subsection{Going from NLO to NNLO processes}
\label{sec:cancel-nnlo}

At present there exists no complete NNLO calculation of the $e^+e^-\to\mu^+\mu^-\gamma(\gamma)(\gamma)$ process. 
A behaviour similar to NLO is expected, i.e., an overall small effect on the cross section, possibly at the level of a few per mil, and significantly larger contributions from hard radiation, which may affect the fiducial acceptance of the analyses. 

\begin{figure}[tbp]
\centering
\includegraphics[width=0.489\textwidth]{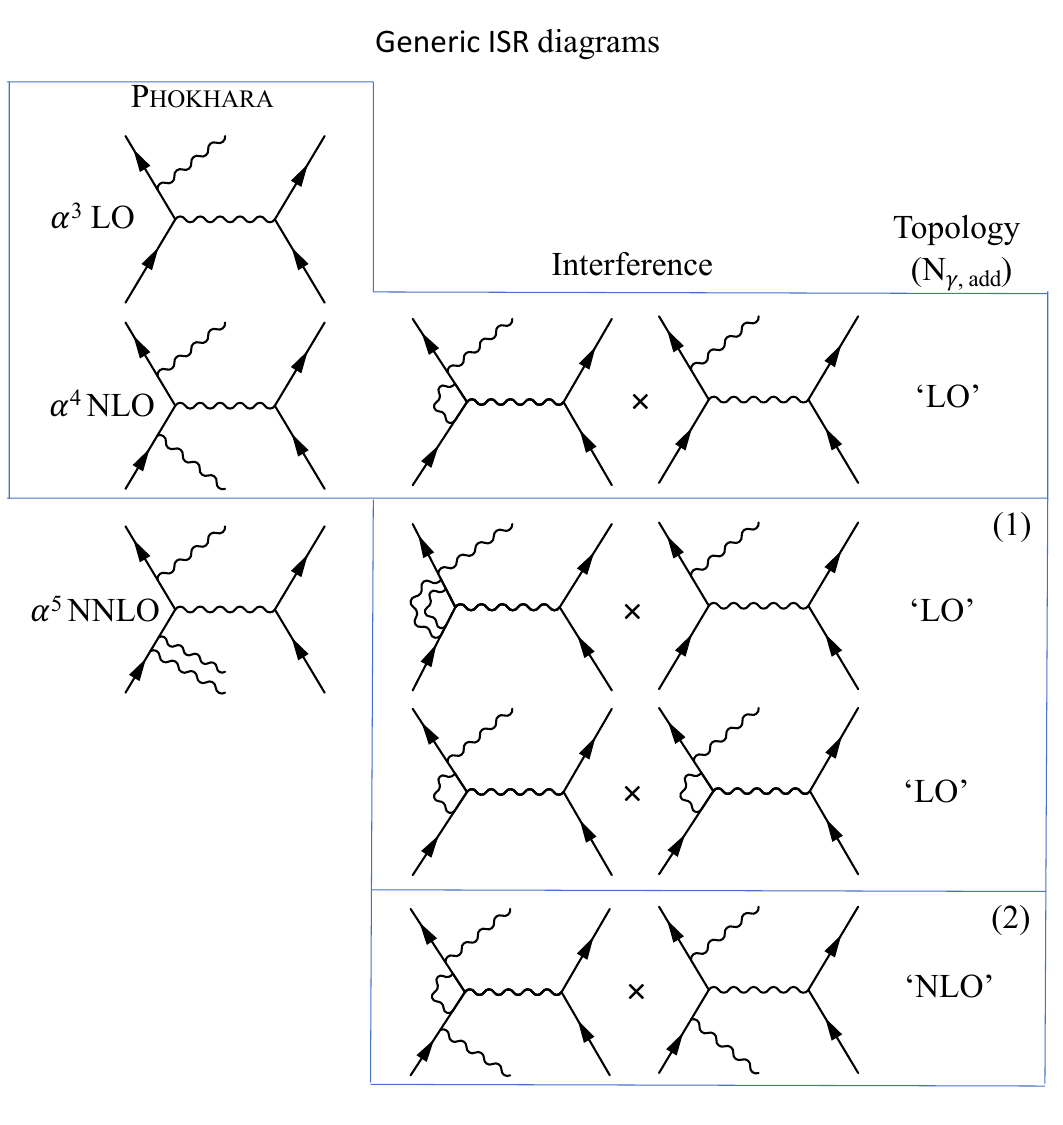}
\caption{\label{fig:feynman} 
Generic Feynman diagrams for the ISR $e^+e^-\to\mu\mu\gamma(\gamma)$ process at LO, NLO and NNLO. At each specified QED order indicated on the left, generic diagrams, ignoring specific topologies with different particle permutations,  are drawn for the virtual (loop) and real photon emission processes. For each order, the latter processes are given on the left-hand side, while the interference contributions are specified on the right together with their experimental topology labeled with quotes. 
$N_\mathrm{\gamma,add}$ refers to the number of real photons emitted, beyond the main ISR photon: $N_\mathrm{\gamma, add} =0$ for the `LO' topology, $N_\mathrm{\gamma, add} =1$ for the `NLO' topology.
In the case of NNLO, the two interference contributions labelled (1) and (2) lead to `LO' and `NLO' topologies, respectively.}
\end{figure}

The investigation of hard and soft/virtual radiative corrections at NNLO is more intricate than at NLO. The situation is illustrated in Fig.~\ref{fig:feynman}, which shows the relevant generic Feynman diagrams. For each order in QED, positive contributions with one to three real photons are separated from contributions from interfering amplitudes involving soft/virtual photons. The first two rows correspond to the diagrams considered in the NLO generator \textsc{Phokhara}. They illustrate the large cancellation occurring at this level as the result of the interference term within the `LO' topology.

At NNLO, the cancellation occurs between the positive three real photon emission contribution and the generic interference contributions leading to an `LO' topology, for the processes labelled (1), or to an `NLO' topology for those in part (2). The interpretation of the results from the radiative BABAR study~\cite{BaBar:2023xiy} depends on the relative importance of these two components. Two extreme scenarios may be considered:
\begin{itemize}
    \item Scenario~1: the processes labelled (1) dominate the NNLO interference term. Since they fall into the `LO' topology, the `NLO' contribution is unaffected by NNLO and the large excess of events predicted by \textsc{Phokhara} compared to the data for the `NLO' topology would need to be interpreted as a generator issue at NLO.
    \item Scenario~2: the processes labelled (2) are the dominant NNLO interference contribution. 
    Being negative, it will affect the `NLO' photon energy distribution in a way uncorrected by the NLO generator. In this situation, the observed deficit in data would arise from NNLO virtual contributions and \textsc{Phokhara} is safe. 
\end{itemize} 
The true situation is likely in-between these two extreme scenarios. Only complete NNLO calculations, at fixed order or in an event generator, will help resolve this ambiguity and should be a high priority for the field. Since the interference contribution listed in the second row of part (1) is obviously positive, it will tend to reduce the negative contribution originating from the first row, perhaps to an overall level smaller than part (2). Also, the interpretation of the BABAR results appears more natural in the second scenario as the NNLO contributions, real and virtual, would explain all the observed features without having to question the validity of the \textsc{Phokhara} generator at NLO.

\section{Impact of higher-order radiative effects}
\label{sec:impact_other}

The observation that \textsc{Phokhara} does not correctly predict the `NLO' contribution raises potential issues for ISR experiments measuring only part of the cross section because of event selection criteria. The fiducial acceptance of an analysis is evaluated with a Monte Carlo generator interfaced with a simulation of the detector response. KLOE, BESIII and CLEOc~\cite{Xiao:2017dqv} rely on \textsc{Phokhara} to estimate the unselected `NLO' part. Hard NNLO contributions are ignored. 
As explained in Section~\ref{sec:cancellations}, a mis-evaluation of hard NLO and NNLO contributions is not compensated by soft/virtual contributions at the same order since the latter are included in the selection of lower-order-like events. 
This unbalance will generate a bias in the cross section measurement. 

\subsection{Procedure}

It is not possible to accurately compute the bias
without full knowledge of the respective analyses and associated detector performance. The purpose of the following study is limited to estimating the possible extent of the bias by reproducing the kinematic conditions of the published analyses  with a simplified generic detector. The study is further restricted to two configurations: `KLOE08' with small-angle undetected ISR~\cite{KLOE:2008fmq} and `BESIII' with large-angle measured ISR~\cite{BESIII:2015equ}, the quotes indicating their generic nature. In both cases, Monte Carlo samples are generated with \textsc{Phokhara} in the kinematic conditions of the experiments with a fast simulation of the tracking and calorimeter performance. 
The 4-vectors from the event generator are converted into pseudo ``reconstructed'' data using the acceptances and resolutions of the detector, as found in the papers published by the experiment. Analysis steps are then applied to these pseudo data to reproduce the overall acceptance and efficiency taking into account the analysis cuts used by the experiment, e.g., fiducial acceptance in the ($M_\mathrm{trk}$,\,$M_{\pi\pi}$) plane for KLOE or $\chi^2_\mathrm{LO}$ selection for BES-III. 

Moreover, two assumptions are made: first, the hard NNLO fraction is taken from the BABAR measurements and assumed to hold independently of the experiment's CM energy. Secondly, real hard NNLO and soft/virtual NNLO radiative corrections are assumed to cancel in the cross section. 
In absence of a complete NNLO calculation, the effect of the observed hard NNLO contribution on the `NLO' spectrum is not known. This ambiguity is related to the relative importance of parts (1) and (2) in Fig.~\ref{fig:feynman}, which will be approached by considering the extreme scenarios introduced above. 

While scenario~2 can be readily transposed to any ISR experiment by estimating the effect of missing NNLO corrections, evaluating the impact of scenario~1 is more delicate without knowing the origin of the issue in the \textsc{Phokhara} generator. It is worth mentioning in this context that all tests documented in the \textsc{Phokhara} publications to evaluate the impact of NLO versus LO corrections relate to the integrated cross section as a function of the two-pion mass~\cite{private-czyz}. Albeit \textsc{Phokhara} was used by the experiments to evaluate the fiducial acceptance and  efficiency of energy and angular selection requirements on additional ISR photons, the modelling accuracy was to our knowledge never tested.  
Experiments exploiting ISR measure the $M_{\pi\pi}$ spectrum of the selected $\pi^+\pi^-\gamma$ sample and correct it for acceptance and selection efficiencies to determine the Born-level $e^+e^-\rar\pi^+\pi^-$ cross section   
\beq
\sigma_{\pi\pi} = \frac{dN_{\pi\pi\gamma}}{dM_{\pi\pi}}\cdot \frac{s}{2M_{\pi\pi}H(M_{\pi\pi})\varepsilon_\mathrm{acc}\varepsilon_\mathrm{sel}L_{ee}}\,,
\eeq
where $s$ is the CM energy squared, $H(M_{\pi\pi})$ the ISR radiation function (radiator), and $L_{ee}$ the $e^+e^-$ luminosity. 
The acceptance $\varepsilon_\mathrm{acc}$, selection efficiency $\varepsilon_\mathrm{sel}$, and $H(M_{\pi\pi})$ are evaluated with \textsc{Phokhara}. Although we have studied NNLO effects for all three variables, results will only be reported for the selection efficiency which is affected most.

\subsection{ Generic `KLOE08' configuration}

In the experimental configuration with $\sqrt{s}=1.02$\gev, only the two charged particles are detected in a polar angle range between $50^\circ$ and $130^\circ$. 
Their three-momenta are measured accurately, while the energy and polar angle of the putative ISR photon are calculated assuming LO kinematics for the ISR process. The ISR photon is required to be emitted in a dead cone of 15$^\circ$ around the beams. The common track mass $M_\mathrm{trk}$ of the two charged particles, computed under the LO assumption, allows to separate dimuon from dipion processes. The selection of $\pi^+ \pi^- \gamma$ events is defined in the ($M_\mathrm{trk}$,\,$M_{\pi\pi}$) plane in a region avoiding background from $\phi\rar\pi^+\pi^-\pi^0$ and the muon band~\cite{KLOE:2008fmq}. 
Fast simulation follows the KLOE performance for charged particle reconstruction~\cite{kloe08-note}. 
Since the selection is very sensitive to the NLO radiative tail, only events satisfying the acceptance cuts are considered here.

\begin{figure*}[tbp]
\centering
\includegraphics[width=\figsize]{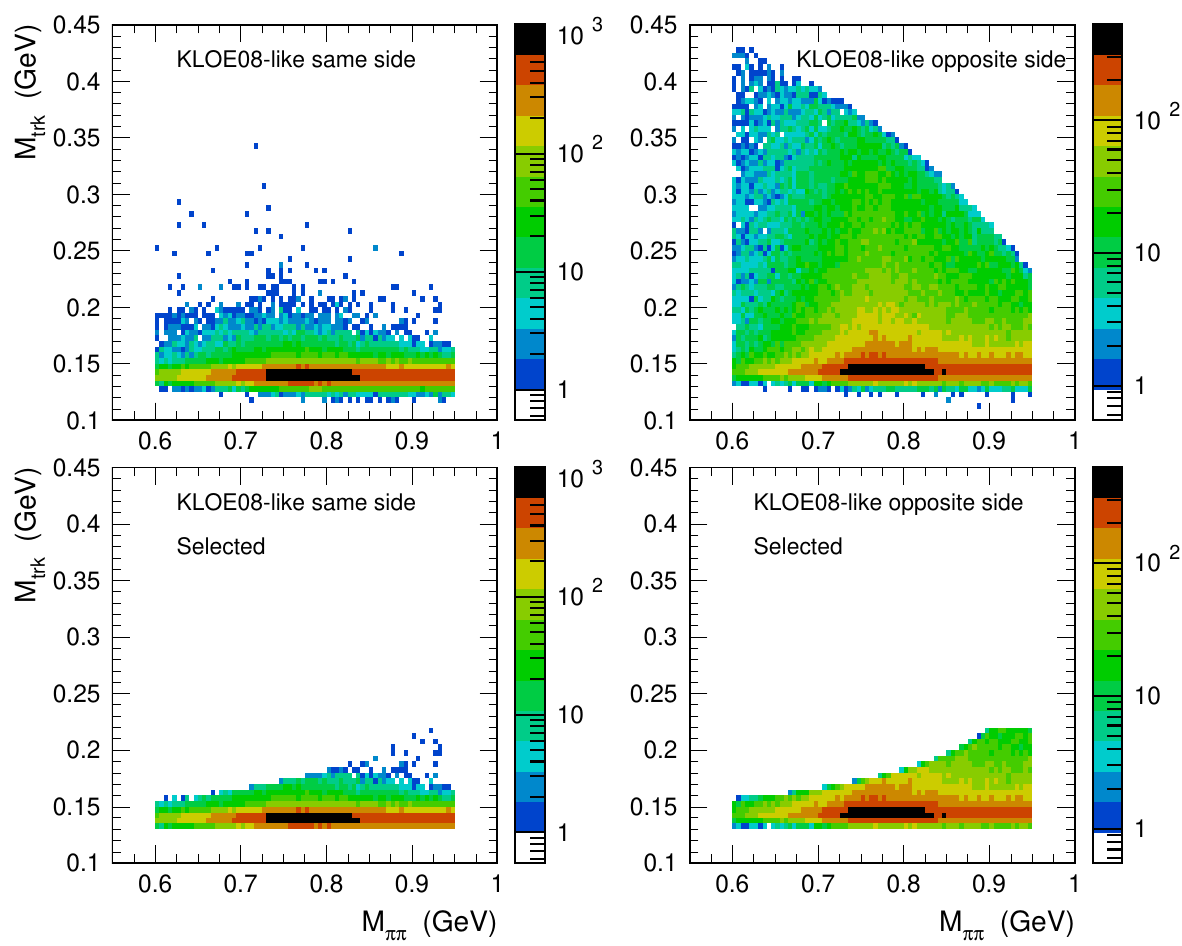}
\vspace{0.2cm}
\caption{\label{fig:mtrk-mpipi} 
Distributions of simulated $e^+e^-\rar\pi^+\pi^-\gamma(\gamma)$ events within the fiducial acceptance for the `KLOE08' configuration in the ($M_\mathrm{trk}$,\,$M_{\pi\pi}$) plane. Top: events with an additional NLO photon  with energy larger than 5\mev emitted in the same hemisphere as the ISR photon (left) and in the opposite hemisphere (right). Bottom: the corresponding distributions after applying selection requirements.}
\vspace{0.4cm}
\centering
\includegraphics[width=\figsize]{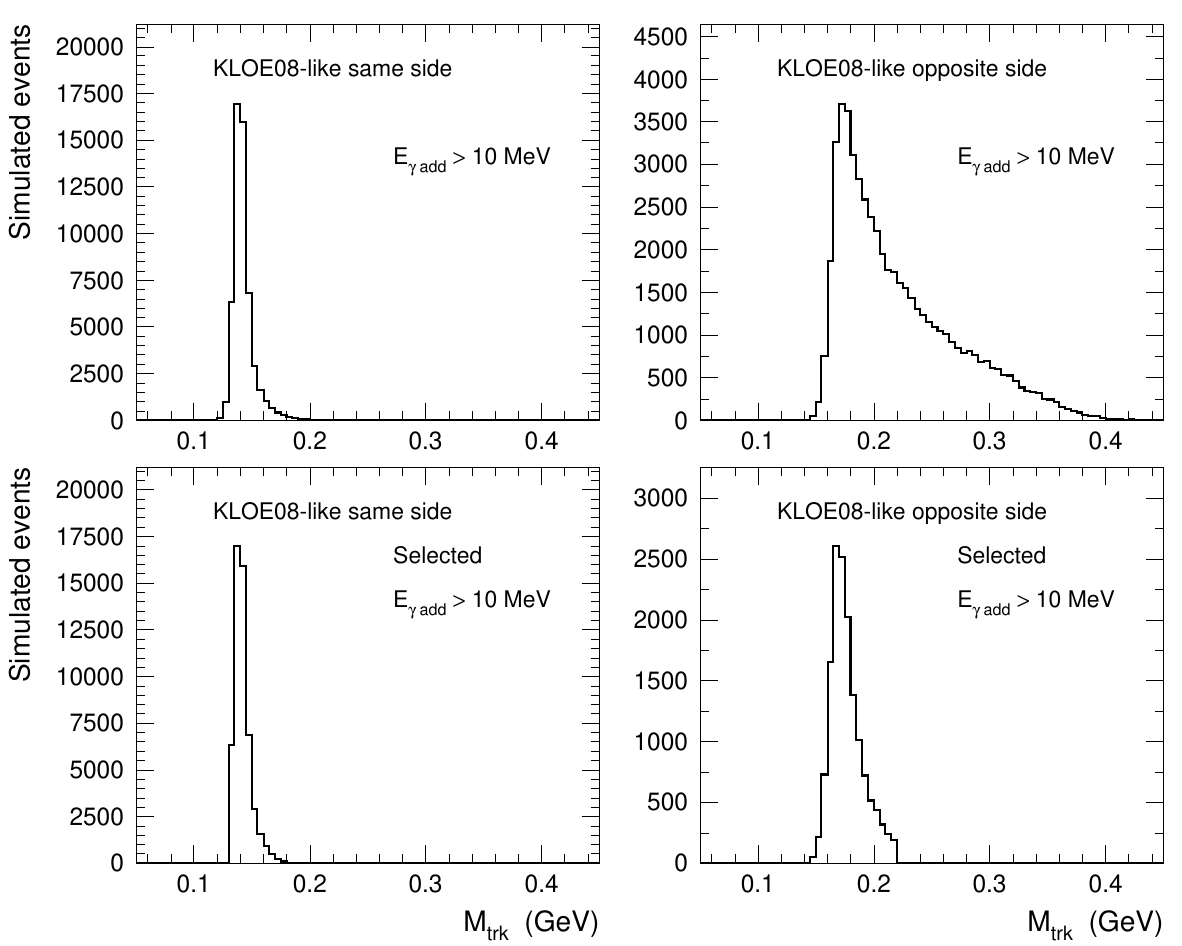}
\vspace{0.2cm}
\caption{\label{fig:mtrk} 
Distributions of $M_\mathrm{trk}$ for simulated $e^+e^-\rar\pi^+\pi^-\gamma(\gamma)$ events within the fiducial acceptance for the `KLOE08' configuration. Top: events with an additional NLO photon with energy larger than 10\mev emitted in the same hemisphere as the ISR photon (left) and in the opposite hemisphere (right). Bottom: the corresponding distributions after applying selection requirements.}
\end{figure*}

Despite the LO-like selection, half of the NLO events are automatically kept in the selected sample when the additional ISR photon is emitted in the same hemisphere as the primary ISR photon, reducing thereby the dependence of the selection efficiency on the \textsc{Phokhara} generator.
This situation occurs since all ISR emissions are sharply peaked in the beam direction, resulting in a small invariant mass of the diphoton system consistent with the zero-mass assumption in the $M_{\rm trk}$ calculation. However, when the additional photon is emitted along the opposite beam direction, the diphoton mass can be relatively large, introducing a long tail in the $M_\mathrm{trk}$ distribution. In that case, the selection efficiency depends on the validity of the photon distribution predicted by \textsc{Phokhara}. The simulated event distributions within the fiducial acceptance in the ($M_\mathrm{trk}$,\,$M_{\pi\pi}$) plane for the  same-side and opposite-side samples are shown for all events (top) and the selected events (bottom) in Fig.~\ref{fig:mtrk-mpipi}. Figure~\ref{fig:mtrk} shows the corresponding $M_\mathrm{trk}$ distributions integrated over $0.6<M_{\pi\pi}<0.95\gev$, for events with additional photon energies larger than 10\mev. Here the opposite side configuration leads to $M_\mathrm{trk}$ values above the pion peak, which are selected with an average efficiency $\varepsilon_\mathrm{sel}^\mathrm{oppo}$ of only 25\%.
 
To estimate the effect of missing hard NNLO radiation in \textsc{Phokhara}, the fraction of $(3.5\pm0.4)$\% observed by BABAR is assumed, as the relative thresholds for the additional photon, 100\mev in BABAR versus 10\mev in KLOE, scaled by the beam energies, are comparable~\footnote{This scaling is approximate because of the phase-space reduction due to the non-negligible muon mass at the low centre-of-mass energy for KLOE.}. Taking further the NNLO contribution as a perturbation of the much larger hard NLO component, the selection efficiency is assumed to be unaffected. Following the previous discussion, 
three out of four configurations feature the emission of at least one of the two additional photons opposite to the ISR photon and thus contribute to the radiative tail of the $M_\mathrm{trk}$ distribution. Averaging over the KLOE08 mass range (dominated by the larger statistics at high mass), the resulting cross section change from the reduced selection efficiency amounts to roughly $-3.5\cdot 3/4 (1 - \varepsilon_\mathrm{sel}^\mathrm{oppo})\% = -2.0$\%. The $M_{\pi\pi}$ dependence is small across the $\rho$ mass region, with a value of $-2.3$\% at the peak. 

In scenario~1 the NLO excess in \textsc{Phokhara} is assumed to be a generator issue. 
Were the NLO fractional excess at the same level as that observed by BABAR, the resulting effect on the selection efficiency would partially cancel the bias from missing NNLO radiation, with a residual effect of order $-1\%$ at the $\rho$ peak.

In scenario~2, the use of \textsc{Phokhara} is safe and the only bias originates from missing NNLO corrections. Hard NNLO radiation contributes as in scenario~1, but its effect is reduced by the negative interference contributions with an `NLO' topology (cf. part (2) in Fig.~\ref{fig:feynman}), of which only one half with opposite-side radiation contributes to the $M_\mathrm{trk}$ radiative tail.  The resulting cross section change is estimated to be $-3.5\cdot (3/4-1/2) (1 - \varepsilon_\mathrm{sel}^\mathrm{oppo})\% = -0.7$\% for the average over $M_{\pi\pi}$, and $-0.8$\% at the $\rho$ peak.

Both scenarios lead to cross section changes that exceed the 0.5\% uncertainty assigned by KLOE08~\cite{KLOE:2008fmq} to radiative corrections.

\subsection{Other KLOE measurements}

The `KLOE10' configuration with the ISR photon detected at large angle and the two pions in the same range as `KLOE08' may be treated in a similar way. Because additional ISR photons predominantly emitted along the beams are well separated from the detected ISR photon, one expects both same and opposite sides to contribute to the $M_\mathrm{trk}$ radiative tail. The cross-section change in scenario~1 is therefore expected to be larger than for the `KLOE08' configuration. In scenario~2, however, since the NNLO positive real photon and negative virtual/soft interference contributions approximately cancel, and lead to photon topologies in the rejected radiative tail, there is no bias for KLOE10.

In the KLOE12 measurement~\cite{KLOE:2012anl} the  cross section was directly obtained from the ratio of the $\pi^+\pi^-\gamma$ to $\mu^+\mu^-\gamma$ mass spectra, protecting the result against modelling biases.
However, in practice, the protection is incomplete as pion and muon selection requirements differ in the ($M_\mathrm{trk}$,\,$M_{\pi\pi/\mu\mu}$) plane. 
While part of the pion radiative tail is retained, it is almost entirely removed in the selected muon sample by a tight $80<M_\mathrm{trk}<115$\mev requirement applied to reduce the pion background. 
This asymmetry in the selection of the pion and muon samples reintroduces a modelling dependence.

KLOE12 features a comparison of the measured muon ISR cross section with the QED NLO prediction by \textsc{Phokhara}. The results show agreement within the quoted systematic uncertainty of 1\%, which is however insufficient to validate the 0.5\% uncertainty assigned to radiative corrections in the two-pion cross-section measurement. A newer dimuon study based on a much larger data set does not improve in precision~\cite{KLOE-2:2016mgi}. We may proceed as in the pion case to estimate the effect of missing NNLO corrections in \textsc{Phokhara}. The cross section change under scenario~1 is found to be of order $-2.6$\%, while the increase from a potential \textsc{Phokhara} NLO excess cannot be estimated. As for the pions, an NLO excess as the one observed in BABAR would essentially cancel that cross-section change. In scenario~2, the muon cross-section change is found to be reduced, as for the pions, by a factor 1/3 to $-0.9$\%. Such a bias would amount to twice the quoted radiative correction uncertainty, but would not be detectable given the 1\% systematic uncertainty of the test.

\subsection{Generic `BESIII' configuration}

\begin{figure*}
\centering
\includegraphics[width=\figsize]{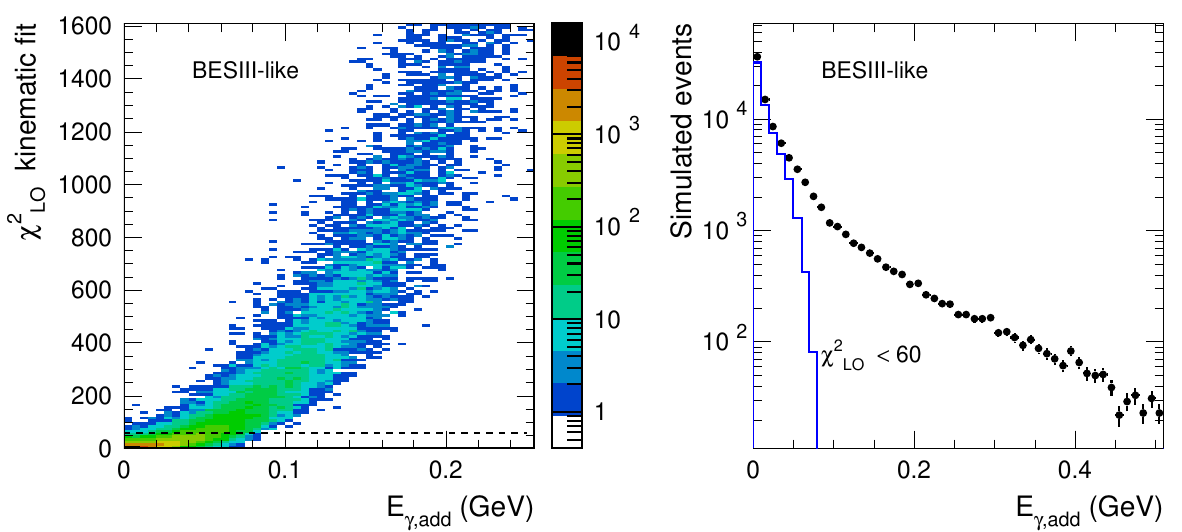}
\vspace{0.2cm}
\caption{\label{fig:chi2-Eadd} 
Distributions of simulated $e^+e^-\rar\pi^+\pi^-\gamma(\gamma)$ events in the `BESIII' configuration. Left: $\chi^2_\mathrm{LO}$ of the $\pi^+\pi^-\gamma$ kinematic fit as a function of the true energy of the NLO additional photon. 
The dashed horizontal line indicates the  $\chi^2_\mathrm{LO}<60$ selection requirement. 
Right: the energy spectrum of the additional NLO photon with (blue line) and without (dots) applying the compatibility requirement.}
\vspace{0.4cm}
\centering
 \includegraphics[width=15cm]{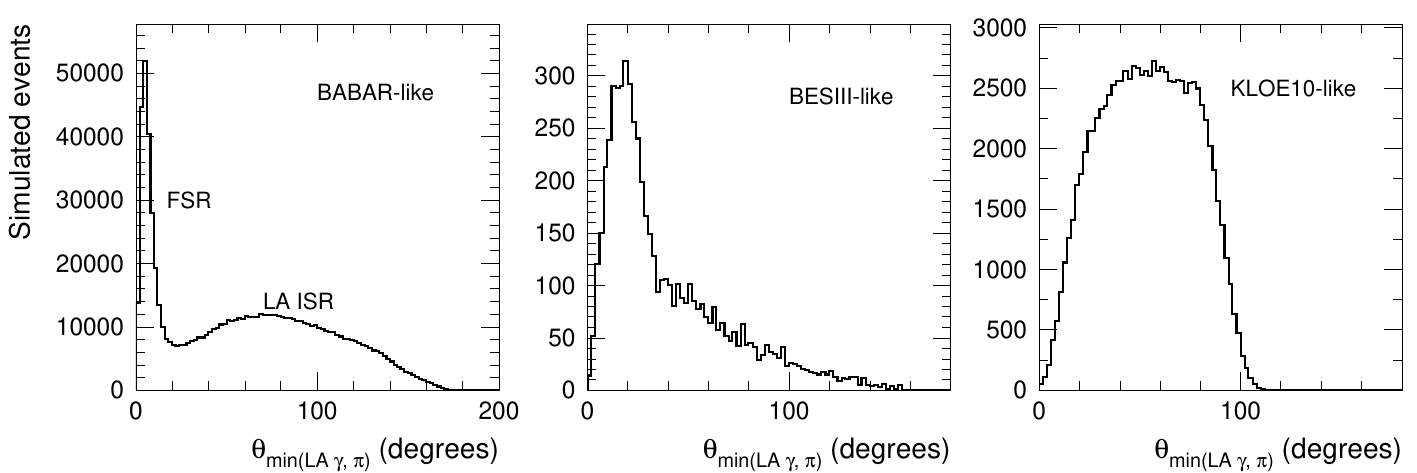}
\vspace{0.2cm}
\caption{\label{fig:isr-fsr-angle} 
The minimum angle between the additional large-angle (LA) photon and the two pions within the detector acceptance for simulated $e^+e^-\rar\pi^+\pi^-\gamma(\gamma)$ events in the BABAR, BESIII, and KLOE10 conditions (left to right panels). 
The separation between FSR and LA ISR events is pronounced at high CM energy (BABAR), still visible at  intermediate CM energy (BESIII), and vanishes at low CM energy (KLOE). }
\end{figure*}

BESIII reported ISR based $e^+e^-\to\pi^+\pi^-\gamma$ cross-section results~\cite{BESIII:2015equ} using data taken at $\sqrt{s}=3.773$\gev, a factor of three below (above) the BABAR (KLOE) CM energy. The analysis requires detection of the two pions and a large-angle ISR photon, while additional photons are ignored. A kinematic fit using the $\pi^+\pi^-\gamma$ hypothesis selects LO and NLO soft/virtual events with the requirement $\chi^2_\mathrm{LO}<60$. A fast simulation of the `BESIII' configuration and detector performance~\cite{changzheng-liangliang}, using the \textsc{Phokhara} generator and the same assumptions as in the `KLOE08' study, allows to 
investigate the effects of additional photon radiation. 

Figure~\ref{fig:chi2-Eadd} (left) shows the distribution of $\chi^2_\mathrm{LO}$ as a function of the true NLO additional photon energy $E_{\gamma,\mathrm{add}}$, exhibiting a strong correlation. 
Events with about $E_{\gamma,\mathrm{add}}> 50\mev$ are subject to rejection. As this maximum accepted $E_{\gamma,\mathrm{add}}$ is consistent with the BABAR threshold of $100$--$200\mev$ for NLO/NNLO photons when normalized to the respective beam energies, one expects very low selection efficiencies of NLO/NNLO events in BESIII.
The distributions of $E_{\gamma,\mathrm{add}}$ for all radiative events and after the $\chi^2_\mathrm{LO}<60$ selection is shown on the right panel of Fig.~\ref{fig:chi2-Eadd}. The fraction of rejected events with $E_{\gamma,\mathrm{add}}>50\mev$ is 92\%. 

The fractional cross-section change due to missing NNLO in \textsc{Phokhara} in scenario~1 amounts to approximately $-3.5\cdot 0.92$\% = $-3.2$\%, again significantly exceeding the assigned systematic uncertainty of 0.5\%. As in the the case of KLOE, this large effect might be partially cancelled by an NLO excess in \textsc{Phokhara} under scenario~1 that we are unable to propagate to CM energies lower than BABAR. 

Similarly to KLOE10, the tight `LO' selected topology preserves BESIII from any bias under scenario~2 as the NNLO positive and negative contributions approximately cancel in the rejected radiative tail.

\subsection{Additional remarks}

The quantitative effects of higher order radiative corrections on the KLOE and BESIII two-pion cross-section results estimated here cannot be taken at face value. Rather, they indicate the potential size of systematic effects encountered from the use of \textsc{Phokhara} in view of the findings reported by BABAR~\cite{BaBar:2023xiy}. 
According to our study, effects from neglected NNLO contributions may suggest upward cross-section corrections that exceed the quoted systematic uncertainties, potentially reducing the difference seen with  BABAR.  
The concomitant effect of the hard NLO excess in \textsc{Phokhara}~\cite{BaBar:2023xiy} is more speculative and may depend on CM energy. Investigations by the generator authors should allow to shed light on this issue~\cite{czyz}. Any definitive assessment needs to be carried out by the KLOE and BESIII collaborations with the full machinery of their analyses.

In this context we also performed a test comparing dimuon samples generated with \textsc{Phokhara} and \textsc{KKMC}~\cite{Jadach:1999vf} in the `KLOE08' configuration.  Differences in the energy distributions of additional photons at `NLO' level lead to different acceptance predictions among the two generators. By construction, \textsc{KKMC} produces higher photon multiplicities, but it predicts the fraction of three or more photons above 10 MeV at KLOE  energies to be 1.3\%, which is lower than the corresponding `NNLO' rate found by BABAR. Of course the two generators operate in different ways. Whereas \textsc{Phokhara} is designed for ISR with an NLO matrix element, \textsc{KKMC} works from the Born level up with multiple ISR photon emission approximating higher orders. It is beyond the scope of this paper to conduct a detailed  evaluation of these generators, but we note their different predictions. Contrary to our above estimates for \textsc{Phokhara}, \textsc{KKMC} would predict a downward shift of the measured cross sections, albeit again larger than the quoted systematic uncertainty assigned to radiative corrections. 

In a recent paper~\cite{Belle-II:2024msd}, Belle II confirms the BABAR finding of an `NLO' excess by \textsc{Phokhara} with respect to their $\pi^+\pi^-\pi^0$ data. To account for this excess and the missing NNLO contributions, they assign a 1.2\% systematic error for the generator.
It is critical that future (re-)analyses of ISR based cross section measurements perform data-driven tests of the kinematic properties of additional photons
as Belle II has done. Such tests allow to investigate the sensitivity to mismodelling and higher order radiative effects, and help design robust selection criteria. 
The loose selection used by BABAR could be implemented rather straightforwardly in the BESIII analysis since the setup follows the same topology with a large-angle ISR photon, and the detector allows the measurement of large-angle additional photons. The situation is more complicated for KLOE as the selection method, at least in the small-angle ISR topology used in KLOE08 and KLOE12, lacks kinematic constraints, preventing the reconstruction of additional small-angle photons. Such an approach would be possible in the KLOE10 topology, but would require independent charged particle identification. 
Another difficulty for KLOE arises from the low centre-of-mass energy and the proximity of the $\rho$ resonance leading to low ISR photon energies that are not as well separated from additional photons as in the case of BESIII and BABAR. This also presents an obstacle to the experimental separation of additional large-angle ISR and FSR photons as was done by BABAR and would be possible with BESIII, as seen from Fig.~\ref{fig:isr-fsr-angle}. 

\section{Reappraisal of $\tau$ spectral functions}
\label{tau-2pi} 

Spectral functions derived from measurements of mass spectra in hadronic $\tau$ decays provide a complementary input, under isospin symmetry and accounting for isospin-breaking corrections, to compute HVP integrals~\cite{Alemany:1997tn}. 
In the late 1990s, thanks to LEP experiments (particularly ALEPH), $\tau$ spectral functions in the two-pion channel were more precise than the available $e^+e^-$ cross sections. In the following decade, both $\tau$ and $e^+e^-$ data were therefore used by our group~\cite{Davier:1997vd,Davier:1998si,Davier:2002dy,Davier:2003pw,Davier:2010fmf,Davier:2010nc,Davier:2013sfa}. 

The  $\tau$ spectral function $v_{\pi^-\pi^0}(s)$ in the $\pi^-\pi^0$ channel is defined by
\beqn
\label{eq:sf}
   v_{\pi^-\pi^0}(s)
   &=&
           \frac{m_\tau^2}{6\,|V_{ud}|^2}\,
              \frac{B_{\pi^-\pi^0}}
                   {B_{e}}\,
              \frac{1}{N_{\pi^-\pi^0}}\frac{d N_{\pi^-\pi^0}}{ds}  \\
   & & 
              \times\,
              \left(1-\frac{s}{m_\tau^2}\right)^{\!\!-2}\!
                     \left(1+\frac{2s}{m_\tau^2}\right)^{\!\!-1}
              \frac{R_\mathrm{IB}(s)}{S_\mathrm{EW}} \,, \nonumber
\eeqn
with
\begin{equation}
\label{eq:rib}
R_{\rm IB}(s)=\frac{\mathrm{FSR}(s)}{G_\mathrm{EM}(s)}
              \frac{\beta^3_0(s)}{\beta^3_-(s)}
              \left|\frac{F_0(s)}{F_-(s)}\right|^2\,,
\end{equation}
and where  $(1/N_{\pi^-\pi^0})dN_{\pi^-\pi^0}/ds$ is the normalised invariant mass-squared ($s$) 
spectrum of the $\pi^-\pi^0$ final state obtained from the combination of spectra from several experiments, $B_{\pi^-\pi^0}$ ($B_{e}$) are the corresponding $\tau$ branching fractions (final state photon radiation implied), and 
$S_\mathrm{EW}$ is an electroweak radiative correction. The $s$-dependent isospin-breaking (IB) corrections are included in  $R_\mathrm{IB}(s)$.
In Eq.~(\ref{eq:rib}), $\beta_{0,-}$ denote 
the pion velocities in the two-pion CM system for the $\pi^+\pi^-$ and 
$\pi^-\pi^0$ final states, respectively.  
$G_\mathrm{EM}(s)$ is the radiative function, correcting from the $\pi^-\pi^0(\gamma)$ to the 
$\pi^+\pi^-$ final states, requiring the addition of the specific FSR contribution to the neutral case. Several model-dependent approaches exist for the small long-distance radiative correction $G_\mathrm{EM}(s)$. The pioneering work of Cirigliano-Ecker-Neufeld~\cite{Cirigliano:2001er} used Chiral Perturbation Theory (ChPT), while vector dominance was the basis of further work by Lopez Castro \textit{et al.}~\cite{Flores-Baez:2006yiq}. The two methods have been known to be in good agreement. More recently, other studies extended the order in ChPT while satisfying short-distance constraints~\cite{Miranda:2020wdg}. Additional  free parameters, however, deteriorate the precision of the prediction. In the longer term, lattice QCD based estimates are expected to become available~\cite{bruno-bern} and will provide an important cross check. 
The form factor ratio ${F_0}/{F_-}$ takes into account the different masses and widths of the charged and neutral $\rho$ mesons and the $\rho$\,--\,$\omega$ interference only present in the neutral final state.

The idea of significant $\rho$\,--\,$\gamma$ mixing, motivated by the well-founded $Z$\,--\,$\gamma$ mixing in high-energy $e^+e^-$ collisions, was put forward by F.~Jegerlehner and R.~Szafron~\cite{Jegerlehner:2011ti} and introduced large IB corrections on top of what had been previously estimated. However, a  justification for applying the same $Z$\,--\,$\gamma$ formalism to the composite $\rho$ meson was never given. It was exacerbated by the proposal by F.~Jegerlehner to reverse the correction and apply it to $e^+e^-$ rather than $\tau$ data~\cite{reverse-rho-gamma}. 
In a consistent dispersive approach of the pion form factor there is no room for $\rho$\,--\,$\gamma$ mixing as differences between charged and neutral $\rho$ line shapes are embedded in their respective resonance parameters (mass, width)~\cite{private-colangelo}. 
The consideration of $\rho$\,--\,$\gamma$ mixing is therefore dropped. 

The use of $\tau$ spectral functions was at some point discontinued owing to the improved $e^+e^-$ cross-section data from KLOE and BABAR not requiring IB corrections. Given the discrepancies among the $e^+e^-$ data sets and the progress on the understanding of IB corrections, we reconsider them here
and present an update of the $2\pi$ HVP contribution to the muon $g$\,--\,2 from $\tau$ decays. The combined $\tau$ mass spectrum, after an update of the ALEPH data, is unchanged from Ref.~\cite{Davier:2013sfa}. A small change is introduced by updated IB corrections,
essentially the $\rho$\,--$\omega$ contribution.
The parameters used in Eq.~(\ref{eq:sf}) are
$m_\tau=(1776.84\pm 0.17)$\mev, the CKM matrix element 
$|V_{ud}|=0.97418\pm0.00019$, and $B_{e}=(17.818 \pm 0.032)\%$. Short-distance electroweak radiative effects~\cite{marciano,Braaten:1990ef}, relevant for the $\pi\pi$ decay give
$S_\mathrm{EW}=1.0235\pm0.0003$~\cite{Davier:2002dy}. 

Most corrections to the $\tau$-based $2\pi$ contribution to $a_\mu$ are unchanged from our previous work~\cite{Davier:2010fmf,Davier:2013sfa}. 
They amount to (all in $10^{-10}$ units): 
$-12.21\pm0.15$ from $S_\mathrm{EW}$,
$-1.92\pm0.90$ from $G_\mathrm{EM}$, 
$+4.67\pm0.47$ from FSR, 
$-7.88$ from $m_{\pi^-}-m_{\pi^0}$ in the cross section, 
$+4.09$ from $m_{\pi^-}-m_{\pi^0}$ in $\Gamma_\rho$, $+0.20^{+0.27}_{-0.19}$ from $m_{\rho^-}-m_{\rho^0}$, 
$-5.91\pm0.59$ from $\pi\pi\gamma$ and other electromagnetic $\rho$ decays. 
The last four corrections are affected by a systematic uncertainty from the choice of the analytic model for the $\rho$ lineshape, which we estimate from the difference between the Gounaris-Sakurai and K\"uhn-Santamaria resonance parameterisations and add linearly.

Due to its fast bipolar dependence on mass the contribution of $\rho$\,--\,$\omega$ interference to the dispersion integral is relatively small. It depends on the $\omega$ mass, the mixing amplitude $\varepsilon_{\rho\omega}$ and its phase $\phi_{\rho\omega}$, all determined from fits to the pion form factor in $e^+e^-$ data. The value for $\phi_{\rho\omega}$ used in our previous analyses~\cite{Davier:2010fmf,Davier:2013sfa} was unexpectedly large~\cite{Colangelo:2022prz}. 
Here, we use updated results from a fit to the combined $e^+e^-$ data before \cmd3~\cite{Colangelo:2020lcg,stoffer-bern} giving $m_\omega=782.07\pm0.15$\mev, $\varepsilon_{\rho\omega}=(1.99\pm0.03)\times 10^{-3}$, and $\phi_{\rho\omega}=(3.8\pm1.8)^\circ$. Including \cmd3~\cite{Colangelo:2023rqr,stoffer-bern} gives similar results with the full difference added as systematic uncertainty. 
The resulting IB correction from $\rho$\,--\,$\omega$ mixing is $+(4.0\pm0.4)\times 10^{-10}$.

Summing up all the effects, the total IB correction to the $\tau$-based $2\pi$ contribution is estimated to be $-(14.9\pm1.9)\times 10^{-10}$ to be compared to our previous estimate of $-(16.1\pm1.9)\times 10^{-10}$~\cite{Davier:2010fmf,Davier:2013sfa}.
Finally the contribution to $a_\mu$ from the combined $\tau$ data reads
\beq
\label{eq:amutau2pi}
 a_\mu^\tau [2\pi] = (517.3 \pm 1.9 \pm 2.2 \pm 1.9)\times 10^{-10}\,,
\eeq
where the uncertainties are from the combined mass spectrum, the branching fractions, and the IB corrections, respectively.

The result (\ref{eq:amutau2pi}) differs from that obtained in Ref.~\cite{Miranda:2020wdg}, $(519.6 \pm 2.8[\mathrm{exp}] ^{+1.9} _{-2.1} [\mathrm{IB}])\times 10^{-10}$ using ${\cal O}(p^4)$ ChPT. Most of the difference is accounted for by their $S_\mathrm{EW}$ value (1.0201), which does not take into account double counting between $S_\mathrm{EW}$ and $G_\mathrm{EM}$ for the subleading non-logarithmic short-distance correction for quarks. 
This effect is responsible for a shift of $1.7\times 10^{-10}$ in $a_\mu^\tau[2\pi]$. 
The remaining difference\footnote{Larger differences are seen when comparing results from individual experiments.} ($0.6\times 10^{-10}$) originates mostly from the $\rho$ width corrections in the pion form factor.

\section{A new perspective on the muon $g$\,--\,2 HVP contribution from the dispersive method}

Having discussed the tensions among the \eetopp cross-section measurements and their possible origins, and reappraised the use of the complementary $\tau$ spectral functions, we proceed with a quantitative study of the dominant HVP contributions to $a_\mu$. 
We consider here only the most precise results. We do not include the CMD-2 measurements~\cite{CMD-2:2003gqi,CMD-2:2006gxt}, whose discrepancy with \cmd3\ is currently under investigation~\cite{logashenko-bern}, and the SND results, which are in a state of flux from the older~\cite{Achasov:2006vp} to the new measurements~\cite{SND:2020nwa} that are still being updated~\cite{kupich-bern}.

For the following exercise, we consider the LO HVP contributions from the $\pi^+\pi^-$ channel in the wide mass range from threshold to 1.8\gev for each experiment. BABAR and the $\tau$ spectral functions extend over the entire interval, while the other experiments cover a more restricted range and are completed near threshold and at large mass with the combination discussed in Section~\ref{sec:tensions-2pi}.
For KLOE, we use the original combined data from Ref.~\cite{KLOE-2:2017fda} and consider two cases: the full available range and a restricted range of 0.6--0.975$\gev$, where the data are most precise and KLOE's weight in the combination is largest (cf. top panel of Fig.~\ref{fig:WeightsChi2}). 
The two-pion contributions are complemented by  the remaining LO HVP, NLO and NNLO HVP, hadronic light-by-light, as well as QED and electroweak contributions, all taken from Ref.~\cite{Aoyama:2020ynm}. The differences in the resulting $a_\mu$ predictions therefore reflect the differences in the two-pion contributions from each experiment, whose uncertainties correspond to the original ones, that is without rescaling to accommodate inconsistencies among data sets. 

\begin{figure}[t]
\centering
\includegraphics[width=0.491\textwidth]{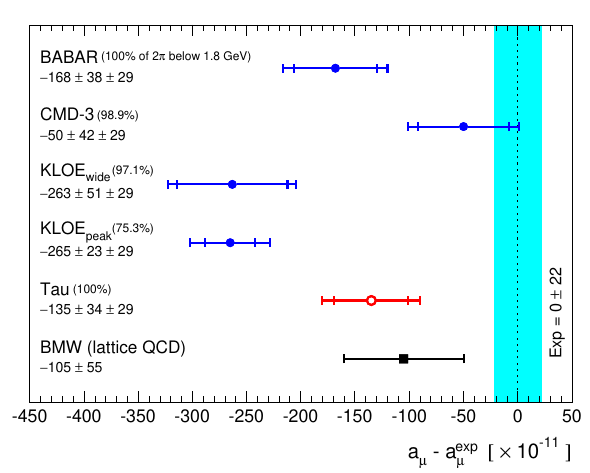}
\vspace{-0.3cm}
\caption{\label{fig:amu-comp} 
Compilation of $a_\mu$ predictions subtracted by the central value of the experimental world average~\cite{Muong-2:2023cdq}. The predictions are computed from the individual \pp contributions between threshold and 1.8\gev, complemented by common non-\pp contributions taken from Ref.~\cite{Aoyama:2020ynm} (circles). The quoted uncertainties correspond to the two contributions and do not include that of the subtracted experimental value shown by the vertical band. The error bars indicate the \pp and total uncertainties, respectively.
The percentage given for each experiment represents the fraction of $a_\mu$[\pp, threshold--1.8\gev] used from a given experiment (see text for details, particularly concerning the two values for KLOE). The lattice result from BMW~\cite{Borsanyi:2020mff} is shown as filled square. }
\end{figure}
\begin{figure}[tbp]
\centering
\includegraphics[width=0.491\textwidth]{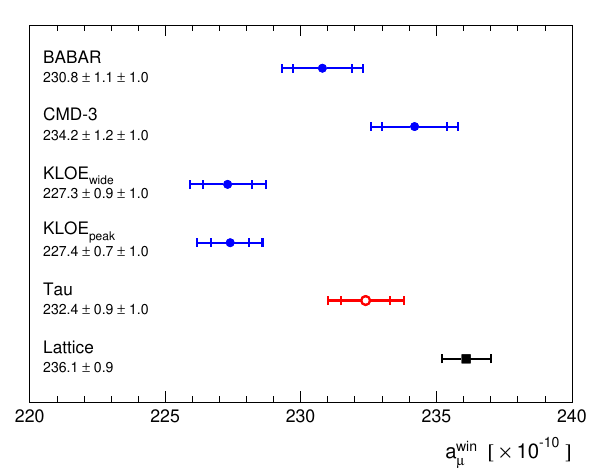}
\vspace{-0.3cm}
\caption{\label{fig:amuWin} 
Compilation of LO HVP $a_\mu^{\rm win}$ predictions in the intermediate Euclidean time window (0.4\,--\,1.0 fm)~\cite{RBC:2018dos}, computed from the individual \pp measurements between threshold and 1.8\gev (when only part of this interval is available it is extended to the full range using Ref.~\cite{Davier:2019can}), complemented by non-\pp combined spectra taken from Ref.~\cite{Davier:2019can}. Also shown is the average of the available lattice QCD results~\cite{Davier:2023cyp}. }
\end{figure}

The results are shown in Fig.~\ref{fig:amu-comp} 
as differences between the  $a_\mu$ predictions and experiment~\cite{Muong-2:2023cdq}. The uncertainties drawn are from the \pp measurements (inner bars) and the total contributions (outer bars). The quoted uncertainties are separated into the \pp and remaining non-\pp contributions. 

The BABAR and $\tau$ based results are in agreement. Combining both with \cmd3 gives $a_\mu^\mathrm{had,\,LO}=(7057\,\pm\,33\,\pm 22)\times 10^{-11}$,
where the first uncertainty is from the \pp contribution, scaled by a factor 1.5 according to the $\chi^2$ value of 4.5 for 2 degrees of freedom and the second from the non-\pp contribution. This average results into
$\Delta a_\mu=a_\mu^{\rm SM}-a_\mu^{\rm exp}=-(123\,\pm\,33\,\pm\,29\,\pm\,22)\times 10^{-11}$, where the first uncertainty is from the \pp contribution, the second from all the other terms in the $a_\mu$ prediction, and the third from the $g$\,--\,2 experimental world average~\cite{Muong-2:2023cdq}. The significance of a non-zero $\Delta a_\mu$ is 2.5$\sigma$. 
As expected from the known tensions, the $a_\mu$ value for KLOE in the restricted range lies well below (3.8$\sigma$) the above combination. 

The BABAR, $\tau$, \cmd3 combination agrees with the only result available so far from lattice QCD for the full $a_\mu$ prediction, BMW~\cite{Borsanyi:2020mff}, who find $\Delta a_\mu=-(105\pm55\pm22)\times 10^{-11}$, shedding a new light on the apparent discrepancy between BMW and the dispersive approach.
Combining the values of BABAR, $\tau$, \cmd3 and BMW, the difference with experiment is $2.8\sigma$.

In the light of these results, we extend the study to the intermediate window 0.4\,--\,1.0\;fm in Euclidean time, which is favourable for lattice QCD. The corresponding $a_\mu^\mathrm{win}$ values are displayed in Fig.~\ref{fig:amuWin}, where the quoted uncertainties are again separated into \pp and non-\pp contributions, the latter contribution using the combined spectra from Ref.~\cite{Davier:2019can}.
\footnote{The $\tau$ based $a_\mu^\mathrm{win}$ result
differs strongly from those given in the first versions of Ref.~\cite{Masjuan:2023qsp}, particularly when using a non-\pp contribution derived from the full spectrum. Because of the different weighting of the mass spectrum, the \pp fraction of the HVP contribution is different for the window (0.64) and the full range (0.73), thus invalidating their procedure. This inconsistency has been corrected in the published version of Ref.~\cite{Masjuan:2023qsp} now in fair agreement with our result.} 
All dispersive predictions are found below that from lattice QCD with significance of $1.1\sigma$ for \cmd3, $2.5\sigma$ for $\tau$, $3.1\sigma$ for BABAR, $5.4\sigma$ for full KLOE, and $5.8\sigma$ for  restricted-range KLOE, exacerbating the pattern seen for $a_\mu$.
The weighted average of  BABAR, CMD-3  and $\tau$ gives $232.0 \pm 1.1$, to be compared with $236.1 \pm 0.9$ from lattice QCD.
To further understand the discrepancy, additional lattice QCD studies, splitting the range of the lattice window into smaller intervals, possibly around the present optimal window, could be helpful.\footnote{See also the discussion on the limitations of such splitting in Ref.~\cite{Davier:2023cyp}.}

\section{Conclusions}

This paper reviewed existing tensions among the most precise \eetopp cross-section measurements used in the dispersive evaluation of the hadronic vacuum polarization (HVP) contribution to the anomalous  magnetic moment of the muon. Local discrepancies between KLOE on one hand and BABAR and \cmd3 on the other hand exceed significances of $3\sigma$ and $5\sigma$, respectively, while that between BABAR and \cmd3 is generally at the $2\sigma$ level. \cmd3 data lie systematically above all other data, while KLOE data lie below. 

A dedicated analysis of radiative processes in $e^+e^- \to \mu^+\mu^-\gamma$ and $e^+e^- \to \pi^+\pi^-\gamma$ at NLO and NNLO by BABAR~\cite{BaBar:2023xiy} prompted a study of 
related systematic uncertainties in the measurements using initial state photon radiation. In absence of an NNLO Monte Carlo generator the studies relied on approximate assumptions and fast simulation. They indicate  potential problems for radiative event acceptances in the KLOE and BESIII measurements, not covered by the quoted systematic uncertainties.  

In view of these difficulties with $e^+e^-$ results we reappraised the use of $\tau$ hadronic spectral functions in the dispersive approach with an updated treatment of isospin-breaking corrections. The $\tau$-based HVP contribution comes out close to the larger values provided by BABAR and \cmd3.

We reevaluated the compatibility of the dispersive HVP calculations with lattice QCD and with the $g$\,--\,2 experiment. Combining  BABAR, $\tau$, and \cmd3 measurements for the \eetopp HVP contribution, and adding all other contributions, the dispersive calculation of $a_\mu$ agrees with the lattice QCD result from BMW~\cite{Borsanyi:2020mff}, while a discrepancy in the restricted observable $a_\mu^{\rm win}$ persists. 

The discrepancy of the dispersive prediction with the $g$\,--\,2 experimental world average reduces from more than $5\sigma$ when KLOE measurements are included but neither CMD-3 nor $\tau$ data, as in \cite{Aoyama:2020ynm}, to the new prediction of $2.5\sigma$ when CMD-3 and $\tau$ measurements are included but not KLOE.

\acknowledgement
\textbf{Acknowledgements} We wish to thank Vincenzo Cirigliano, Gilberto Colangelo, Henryk Czy\.z, Laurent Lellouch, Gabriel Lopez Castro, and Pablo Roig for discussions that were very helpful to clarify various theoretical aspects. We thank Zbigniew W\c{a}s for advice on the KKMC generator, Swagato Banerjee for providing an operational version of the code and Xiaowen Su for generating samples.
This work benefited from funding by the French National Research Agency under contract ANR-22-CE31-0011.


\begin{thebibliography}{99}
\bibitem{czyz}           We shared the information about the found problems with Henryk Czy\.z, one of the \textsc{Phokhara} authors.
\end{thebibliography}
\end{document}